\documentclass[
 pre, nofootinbib
]{revtex4-1}

\usepackage[T1]{fontenc} 
\usepackage{amsmath,amsfonts,amssymb,physics,eqnarray}
\usepackage{graphicx,float,caption,subcaption,xcolor,enumitem}
\usepackage{soul}
\usepackage{dcolumn}
\usepackage{bm}

\begin{document}


\title{Why ETH? On thermalization and locality.}

\author{Stefan Eccles}%
    \email{stefan.eccles@oist.jp}
    \affiliation{Okinawa Institute of Science and Technology Graduate University, Onna, Okinawa 904 0495, Japan}


\begin{abstract}
The eigenstate thermalization hypothesis (ETH) is foundational to modern discussions of thermalization in closed quantum systems.
In this work we expand on traditional explanations for the prevalence of ETH by emphasizing the role of operator locality.  We introduce an operator-specific perturbation problem that can be thought of as a means of understanding the onset or breakdown of ETH for specific classes of operators in a given system.  We derive explicit functional forms for the off-diagonal variances of operator matrix elements for typical local operators under various `scrambling ansatzes', expressed in terms of system parameters and parameters of the corresponding perturbation problem.  We provide simple tests and illustrations of these ideas in chaotic spin chain systems.
\end{abstract}

\maketitle
\tableofcontents

\section{Introduction}
\label{sec:intro}

Any complete understanding of statistical mechanics and thermodynamics must include an account of how, when, and why certain states and observables thermalize while others do not.  In the setting of closed quantum systems, the predominant framework for understanding thermalization is the eigenstate thermalization hypothesis (ETH) \cite{Jensen:PhysRevLett.54.1879,Deutsch:PhysRevA.43.2046,Srednicki:PhysRevE.50.888}. 
Its relevance, and that of its various extensions, has been confirmed numerically in a wide variety of systems \cite{PhysRevE.87.012118,PhysRevE.89.042112,PhysRevE.90.052105,PhysRevE.89.062110,PhysRevE.93.032104,PhysRevE.96.012157,PhysRevB.99.155130,PhysRevE.100.062134, PhysRevLett.125.070605,Noh_2021,Caceres:2024mec, PhysRevB.109.045139,Lin:2024vji,Lasek:2024ess}.  It has been well-motivated from its inception by arguments linking it to quantum chaos and random matrix theory \cite{DAlessio:2015qtq}.  Partial proofs have been given in specific contexts (we mention only a few below).  In light of these facts, the general importance of ETH is not in question; rather, what remains is to delineate its precise domain of applicability: to which operators does it apply, and in which systems?  

In this work we propose a framework for understanding when and why given (sets of) operators are well described by the ETH ansatz for matrix elements by focusing on a general notion of locality in finite systems.  We do so by modifying a classic approach due to Deutsch (\cite{Deutsch:PhysRevA.43.2046, Deutsch:2018ulr}, see also \cite{Reimann_2015}), which is based on perturbing an integrable system to a chaotic system.  Rather than perturbing between two pre-specified Hamiltonians in theory space, we use the locality properties of a given operator or algebra to identify a perturbation problem that is pertinent to the thermalization of those particular operators in a fixed system.  In chaotic systems, the resulting perturbation problem is not amenable to exact techniques, but may motivate various ``scrambling'' ansatzes that determine the extent to which these operator(s) are ETH-class  within a given system.  Such ansatzes can then be used to characterize substructure within, or deviations from, the basic ETH form.  We consider the implications of this line of reasoning for local operators, deriving predictions for the functional form of the off-diagonal ETH ansatz for local operators expressed in terms of system parameters, and parameters of this perturbation problem.  We find good agreement with numerical tests in a chaotic spin chain example.  We also illustrate the use of the framework to understand aspects of matrix element structure in regimes where ETH does not occur.

The paper is organized as follows.  
In section \ref{sec: ETH as hypothesis} we briefly summarize the ETH, framed as an ansatz for operator matrix elements in the energy eigenbasis.  We discuss its status as a `hypothesis' in current research, and summarize the lens through which we view the ansatz in the present work: not as a hypothesis awaiting proof in a various systems, but as a benchmark form for comparison that illuminates the relationship between specific classes of operators and a given system.
In section \ref{sec: where does ETH come from?} we briefly summarize a some standard explanations of the ETH ansatz, in particular an approach pioneered by Deutsch which we later adapt.
In section \ref{sec: ETH through locality} we discuss operator localizability.  We ask the question: given an operator's spectrum, what is the smallest tensor factor, under any tensor product structure (TPS) on the Hilbert space, on which the operator can be considered to act locally.  Given such a factorization, there is an obvious decomposition induced on the system Hamiltonian, and a corresponding perturbation problem, which is pertinent to the thermalization properties of this specific operator.  We discuss the use of this operator-specific perturbation problem to assess the extent to which the operator falls under the ETH framework. 
In section \ref{sec: scrambling ansatzes} we consider some simple ansatzes relating perturbed to unperturbed eigenstates, studying the implications for matrix elements of local operators.  We derive explicit expressions for the decay of the variance of off-diagonal matrix elements for typical local operators in various approximation regimes. 
In section \ref{sec: numerical illustrations}, we provide some numerical illustrations of these results in a chaotic spin chain.
Finally, in section \ref{sec: discussion and future directions} we briefly conclude with some comments and directions for future work.  Certain calculational details, as well as illustrations of the framework in non ETH-like regimes, are placed in appendices.

\section{Eigenstate thermalization as `ansatz' and `hypothesis'}
\label{sec: ETH as hypothesis}
The ETH may be framed in terms of an ansatz for operator matrix elements in the ordered energy eigenbasis \cite{Srednicki_1999}:  
\begin{equation}\label{eq: ETH ansatz}
        \hat{\mathcal{O}}^{(\text{ETH})}_{\alpha\beta}=\mathcal{O}(\bar{E}_{\alpha\beta})\delta_{\alpha\beta} + e^{-S(\bar{E}_{\alpha\beta})/2}f_{\hat{\mathcal{O}}}(\bar{E}_{\alpha\beta},\omega_{\alpha\beta})R_{\alpha\beta}.
\end{equation}
We will refer to Eq. \eqref{eq: ETH ansatz} as defining the ``basic ETH'' form.  The $\hat{\mathcal{O}}_{\alpha\beta}:=\bra*{E_\alpha}\hat{\mathcal{O}}\ket*{E_\beta}$ are matrix elements of some quantum operator $\hat{\mathcal{O}}$ between energy basis states, $\hat{H}\ket*{E_\alpha}=E_\alpha\ket*{E_\alpha}$.  The parameters $\bar{E}_{\alpha\beta}$ and $\omega_{\alpha\beta}$ are the average and difference between these energy eigenvalues (we will sometimes leave off the subscripts, but they will be useful in later sections).
\begin{equation}\begin{split}
    \bar{E}_{\alpha\beta}:=\frac{1}{2}(E_\alpha+E_\beta)\,, \qquad
    \omega_{\alpha\beta}:=\frac{1}{2}(E_\alpha-E_\beta)\,.
\end{split}\end{equation}
The key aspects of the ansatz are that the diagonal matrix elements $\mathcal{O}_{\alpha\alpha}$ vary as a smooth function of the energy, $\mathcal{O}(\bar{E})$, while off-diagonal elements fluctuate `randomly' and are exponentially suppressed in the thermodynamic entropy $S(\bar{E})$ (diagonal fluctuations are likewise suppressed).  It may be directly confirmed that such operators  thermalize for a wide variety of initial states. 
See section 4 of \cite{DAlessio:2015qtq} for a nice summary, and see \cite{Caceres:2024mec} for discussion of how far such a statement can be pushed.

Some descriptions of the ETH include only the features mentioned above, but since we will employ ansatz \eqref{eq: ETH ansatz} and put some emphasis on the off-diagonal behavior, we now briefly explain the other quantities appearing in the expression. The $R_{\alpha\beta}$ are elements of a pseudorandom matrix with $\mathcal{O}(1)$ elements, while $f_{\hat{\mathcal{O}}}(\bar{E},\omega)$ is a smooth function.  These can be required to satisfy different criteria based on symmetry properties of the Hamiltonian.\footnote{For instance if the system obeys a time-reversal symmetry, the eigenvectors of $\hat{H}$ and likewise the matrix elements $\mathcal{O}_{\alpha\beta}$ of Hermitian operators may be chosen to be real, so that
$R_{\alpha\beta} = R_{\beta\alpha}$ and $f_{\hat{\mathcal{O}}}(\bar{E},-\omega)= f_{\hat{\mathcal{O}}}(\bar{E},\omega)$. Lacking any such symmetry, we instead have complex elements with $R^*_{\alpha\beta} = R_{\beta\alpha}$ and $f^{*}_{\hat{\mathcal{O}}}(\bar{E},-\omega)= f_{\hat{\mathcal{O}}}(\bar{E},\omega)$. Alternatively we may require $f$ to be real and absorb all phase information into $R_{\alpha\beta}$, as we do in this work.  The matrix elements $R_{\alpha\beta}$ are usually taken to have mean zero and variance (in magnitude) of one, except in the case of time-reversal symmetry when the diagonal elements have variance two.}  In the basic ETH, the $R_{\alpha\beta}$ are often taken to be independent gaussian variables, though such simplifying assumptions are insufficient for some purposes \cite{Foini:2018sdb, Chan:2018fsp, Dymarsky:2018ccu, Belin:2021ryy}.
Likewise, the detailed form of $f_{\hat{\mathcal{O}}}(\bar{E},\omega)$ is context-dependent, expressing physics beyond (or substructure within) the basic ETH framework, relaying information about nonequal-time correlation functions and linear response to perturbations about equilibrium \cite{Srednicki_1999,PhysRevLett.111.050403}, as well as various benchmarks of chaos and other thermodynamic properties (e.g. see \cite{Murthy:2019fgs,Sorokhaibam:2022tgq,Sorokhaibam:2024fcv}).

Beyond stating the ansatz, the `hypothesis' of ETH lies in the assertion that most or all physically-relevant operators will take the ETH form, and hence will thermalize, in a sufficiently chaotic system.\footnote{An alternative formulation states that most or all energy eigenstates, perhaps in a restricted window, behave as the ansatz implies for a fixed class of operators, giving expectation values matching those of a thermal ensemble. The `most' or `all' conditions leads to a distinction between weak and strong ETH (see, e.g. \cite{PhysRevLett.105.250401, PhysRevE.90.052105,Kawamoto:2024vzd}).  In this work we focus on the structure of operator matrix elements.}  Usually the relevant operators are taken to be `local' or `few-body' operators, but it may also be `macroscopically course-grained' or otherwise `simple' operators, suitably defined.  Of course in \textit{any} system, regardless of how chaotic, one may with equal ease write down operators in the energy basis which take the ETH form and those which do not. Together with the fact that most `partial proofs' of ETH\footnote{To list but a few of these: von Neumann's original approach \cite{vonNeumann_1929,von_Neumann_2010} to thermalization has been revitalized \cite{Goldstein_2010} and improved \cite{PhysRevLett.115.010403} in relation to ETH as a form of typicality condition.  \cite{Ishii_2019} proves a version of generalized ETH for translation-invariant noninteracting integrable systems.  \cite{Helbig:2024lvx} demonstrates that ETH applies to observables on small subsystems given only some simple assumptions about the eigenvalue distribution and interactions (identifying relevant criteria in a manner close in spirit to this work).  \cite{Kawamoto:2024vzd} outlines a strategy of proof in holographic systems.} invoke tools of random matrix theory, this indicates that ETH should be understood a kind of operator typicality statement regarding how operators of a specific class relate to the system Hamiltonian.  In fact, it has also been understood \cite{Jafferis:2022uhu,jafferis2023jtgravitymattergeneralized,deBoer:2023vsm} that the ETH emerges naturally from a principle of maximal ignorance \cite{RN7,PhysRev.106.620}.  Seeking a maximal entropy operator ensemble subject to constraints on the thermal one-point function and thermal time-evolved two-point function leads to a version of the ansatz.  It may therefore be thought of as expressing a `most likely' form of an operator subject to the stated constraints without bias by any other conditions.  On the other hand, this does not clarify the physical conditions under which particular operators actually take this form in a given system.  This work focuses on the latter question, introducing a framework for assessing the reason and extent to which given operators approach the ETH form, aiming in particular to provide an intuitive account of the role of locality.  

To understand why in chaotic systems certain operators naturally take the ETH form, it is also fruitful to consider the opposite question: to which systems and operators does ETH \textit{not} apply?   At the extreme end, there are completely integrable, non-interacting, and few-body systems where nothing resembling ETH could be fruitfully invoked.  Then there are intermediate regimes where modified versions of ETH may apply.  Many-body localizing systems violate the strictest tenets of chaos and ETH, exhibiting phases that resist thermalization of local operators \cite{PhysRevLett.111.127201,PhysRevB.90.174202, annurev:/content/journals/10.1146/annurev-conmatphys-031214-014726,RN8,RevModPhys.91.021001}.  In some integrable systems, operators relax to be described by generalized ensembles \cite{Rigol_Dunjko_Yurovsky_PhysRevLett.98.050405}, a behavior which can be understood via generalized eigenstate thermalization \cite{PhysRevLett.106.140405,PhysRevA.87.063637,Vidmar_2016}, in which eigenstate expectation values vary as smooth functions of conserved quantities in addition to energy.   Some systems exhibit chaos except for the presence of a simple symmetry, in which case ETH may apply separately within symmetry sectors (examples can be found in \cite{santos2010onset,Caceres:2024mec}).  Yet other systems contain multiple noncommuting conserved charges, requiring the basic ETH to be modified as has been recently explored \cite{PhysRevLett.130.140402, Lasek:2024ess}.  Even where the basic tenets of ETH do apply, there is a need to supersede the common simplifying assumptions regarding independent, gaussian-distributed matrix elements to accurately capture the physics of OTOCS and higher n-point functions \cite{Foini:2018sdb, Chan:2018fsp, Dymarsky:2018ccu, Belin:2021ryy}. In view of these manifold generalizations and departures from the basic ansatz, one overarching goal of ETH-related research might be considered the classification of its modes of failure, identifyng what physical circumstances or criteria cause the breakdown of standard ETH, and finding useful characterizations of departure from the basic ansatz.   
It is in this spirit that we view ETH in this work, not as a hypothesis awaiting proof in some particular system but as a benchmark form for comparison in the characterization of the relationship between systems and operators.  The operator-specific assessment described in the following sections is intended as a means of appraising the features of the operator-system relationship that imply ETH or its breakdown.

\section{Where does ETH come from?}
\label{sec: where does ETH come from?}
\subsection{Random matrix theory}
\label{subsec: RMT result}
As a first pass at understanding the ETH ansatz, we review a standard connection to random matrix theory by repeating an argument in section 2 of \cite{DAlessio:2015qtq}.  We ask what a fixed operator looks like in a ``typical'' energy eigenbasis, with the notion of typicality set by averaging over Hamiltonians in a random matrix ensemble \cite{Meh2004}. When necessary for concreteness, we will use the Gaussian Unitary Ensemble (GUE) and in some later sections also the Gaussian Orthogonal Ensemble (GOE).  We denote the eigenvalues and eigenvectors of some operator $\hat{\mathcal{O}}$ according to $\hat{\mathcal{O}}\ket*{\mathcal{O}_\lambda}=\mathcal{O}_\lambda \ket*{\mathcal{O}_\lambda}$, while for the Hamiltonian we write $\hat{H}\ket*{E_\alpha} = E_\alpha \ket*{E_\alpha}$.  Both $\alpha$ and $\lambda$ take values from 1 to the dimension of the hilbert space $\mathcal{H}$, which we denote $|\mathcal{H}|$.  Expanding an energy-basis matrix element of $\hat{\mathcal{O}}$ in terms of the operator eigensystem, we have
\begin{equation}\begin{split}\label{eq: rmt setup}
    \mathcal{O}_{\alpha\beta}:=\bra*{E_\alpha}\hat{\mathcal{O}}\ket*{E_\beta}
    =\sum_\lambda \mathcal{O}_\lambda \braket{E_\alpha}{\mathcal{O}_\lambda}\braket{\mathcal{O}_\lambda}{E_\beta}.
\end{split}\end{equation}
At this point, the properties of the GUE are invoked.  In a large Hilbert space, the eigenvectors are effectively random orthonormal vectors in the $\hat{\mathcal{O}}$ eigenbasis, the correlations between which can be ignored.  Their overlap coefficients with the $\hat{\mathcal{O}}$ eigenbasis, $\braket{\mathcal{O}_\lambda}{E_\alpha}$, are then treated as random complex numbers, independent Gaussian variables of mean zero and variance as required by normalization. 
To leading order in $1/|\mathcal{H}|$, this implies an expression for the matrix elements of $\hat{\mathcal{O}}$ in a `typical' eigenbasis (so defined) as
\begin{equation}\label{eq: RMT result}
    \mathcal{O}_{\alpha\beta} = \overline{\mathcal{O}}\delta_{\alpha\beta} + \sqrt{\frac{\overline{\mathcal{O}^2}}{|\mathcal{H}|}}R_{\alpha\beta},
\end{equation}
where $\overline{\mathcal{O}}$ and $\overline{\mathcal{O}^2}$ are the average of the spectrum and squared spectrum of $\hat{\mathcal{O}}$, respectively, and $R_{\alpha\beta}$ is a random matrix with elements of mean zero and variance one.
In equation \eqref{eq: RMT result} we see the crude beginnings of (or a limiting case of) the ETH ansatz \eqref{eq: ETH ansatz}, particularly in the clear differentiation between diagonal and off-diagonal elements.  The impartial ensemble averaging has ``washed out'' all other structure of the operator, where the ETH retains more. The latter is said to reduce to the random matrix result across small energy windows, a point which we will return to later.  To understand the broad-scale features of ETH, such as the smooth diagonal function of the energy and the $\omega$-dependend off-diagonal decay, we must go further.

\subsection{Deutsch's approach}
\label{subsec: Deutschs approach}
We now consider a ``derivation'' of ETH pioneered by Deutsch \cite{Deutsch:PhysRevA.43.2046, Deutsch:2018ulr}. See \cite{Reimann_2015} for a nice summary and extended analysis with this model.  This approach considers an integrable Hamiltonian $\hat{H}^{(0)}$, to which a small integrability-breaking perturbation $\hat{H}^{(I)}$ is added to give the total chaotic Hamiltonian $\hat{H}^{(T)}=\hat{H}^{(0)}+\hat{H}^{(I)}$. The structure of the perturbing Hamiltonian is specified in the eigenbasis of the unperturbed Hamiltonian\footnote{Superscripts in pararenthesis on the energies will indicate which Hamiltonian the eigenstate/eigenenergy corresponds to, for example $E^{(0)}_i$ for the unberturbed $\hat{H}^{(0)}$ eigenbasis and $E^{(T)}_i$ for the total $\hat{H}^{(T)}$ eigenbasis.}: 
\begin{equation}
    h_{ij}:=\bra*{E^{(0)}_i}\hat{H}^{(I)}\ket*{E^{(0)}_j}.
\end{equation}
This is taken to be a banded matrix, often very sparse, with the typical magnitude of nonzero elements decreasing to zero with increasing $|E^{(0)}_i-E^{(0)}_j|$.  The scale of $\hat{H}^{(I)}$ must be small enough that the perturbation does not significantly alter the system's density of states. With this setup in mind, one then considers \textit{random} perturbations of similar structure, with a diagonal matrix ($\hat{H}^{(0)}$) perturbed by a sparse, banded matrix ($\hat{H}^{(I)}$) of random elements of decreasing magnitude away from the diagonal. Limited rigorous results from random matrix theory are available pertaining to precisely this case (see discussion in \cite{Reimann_2015} and references therein), but an important and seemingly robust feature of this type of perturbation problem is that the eigenstates of the total Hamiltonian $\hat{H}^{(T)}:=\hat{H}^{(0)}+\hat{H}^{(I)}$ tend to become scrambled, but still localized, in the unperturbed eigenbasis.  In other words, expressing a total eigenstate $\ket*{E^{(T)}_\alpha}$ in the unperturbed eigenbasis leads to
\begin{equation}\begin{split}
    \ket*{E^{(T)}_\alpha}&=\sum_i c_{\alpha i} \ket*{E^{(0)}_i},
\end{split}\end{equation}
where the $c_{\alpha i}$ describe a `random' superposition of unperturbed eigenstates centered around $E_i^{(0)}=E^{(T)}_\alpha$.  The detailed statistics of the $c_{\alpha i}$ should be inherited from a selected ensemble of perturbations $\hat{H}^{(I)}$, but making simple assumptions about the statistics of the $c_{\alpha i}$ is fruitful:  writing a matrix element of some (any) operator in terms of the unperturbed basis as
\begin{equation}\begin{split}\label{eq: exact Deutsch}
    \bra*{E^{(T)}_\alpha}\hat{\mathcal{O}}\ket*{E^{(T)}_\beta}
    =\sum_{i,j}c^{*}_{\alpha i} c_{\beta j}\bra*{E^{(0)}_i}\hat{\mathcal{O}}\ket*{E^{(0)}_j}\\
\end{split}\end{equation}
and averaging the coefficients $c_{\alpha i}$ over an ensemble that is \textit{now localized to an energy shell} around $E^{(0)}_i\approx E^{(T)}_\alpha$ allows one to confirm the basic properties of the ETH ansatz.  In particular, the diagonal elements now effectively perform a microcanonical average in the old eigenstates, leading to the emergence of a smooth function of energy on the diagonal, plus fluctuations which are exponentially suppressed in the system size.  Off-diagonal terms can likewise be shown to be suppressed and of mean zero, tending toward zero away from the diagonal under plausible assumptions about the scrambled eigenstates \cite{Reimann_2015}.

A major achievement of this model is the intuitive appearance of a smooth function of the energy on the diagonal, which will inevitably correspond to thermal values.  There are at least two puzzling aspects of the model's success, however.  First, it is not always clear how to identify the relevant perturbation problem.  The setup just described is natural when considering quench scenarios, for example, where some initial Hamiltonian $\hat{H}=\hat{H}_{\text{initial}}$ is suddenly altered to a new Hamiltonian $\hat{H}=\hat{H}_{\text{final}}$, in order to study equilibration with respect to the latter. An obvious choice is then $\hat{H}^{(0)}=\hat{H}_{\text{initial}}$ and $\hat{H}^{(I)}=\hat{H}_{\text{final}}-\hat{H}_{\text{initial}}$.  But in principle, the ETH form relates an operator only to the total Hamiltonian $\hat{H}^{(T)}=\hat{H}_{\text{final}}$, so it is unclear why some $\hat{H}^{(0)}$ should feature heavily in the analysis.  Deutsch points out (section IV of \cite{Deutsch:2018ulr}) that there is no reason to require $\hat{H}^{(0)}$ to be integrable.  One supposes that long as $\hat{H}^{(T)}$ is in a chaotic regime, the precise choice of $\hat{H}^{(0)}$ (equivalently, $\hat{H}^{(I)}$) does not matter; what is really leveraged is the expectation that nearby points in theory space lead to a similar localized scrambling of eigenstates.  With this line of reasoning, one must still decide how to parametrize theory space so that the relevant class of perturbations is clear.

A second puzzling feature is that the approach treats all operators on an equal footing.  Viewing ETH as an operator typicality statement, this is not a flaw per se, but one might be led to conclude that \textit{all} operators are equally destined to take the ETH form, which of course cannot literally be true.  Perhaps a more accurate impression is that all operators are equally \textit{likely} to take the ETH form, given only some scrambling statistics of the coefficients $c_{\alpha i}$.  But so far the approach offers no hints as to any distinguishing features among the operators themselves.

In what follows, we suggest a remedy to both of these objections through a simple modification of the procedure that emphasizes the role of locality.

\section{ETH through locality}
\label{sec: ETH through locality}
\subsection{Operator localizeability}
\label{subsec: localizeability}
Let us now clarify the relevant notion of locality.  Suppose we have a finite, bipartite Hilbert space comprised of subsystems $A$ and $B$:
\begin{equation}\label{eq: bipartite hilbert space}
    \mathcal{H}:=\mathcal{H}^{(A)}\otimes\mathcal{H}^{(B)}.
\end{equation}
An operator that is ``local to $A$'' is one which can be written as
\begin{equation}\label{eq: local op}
    \hat{\mathcal{O}}:=\hat{\mathcal{O}}^{(A)}\otimes \mathbb{I}^{(B)},
\end{equation}
where $\mathbb{I}^{(B)}$ is the identity operator on $\mathcal{H}^{(B)}$.
A shared feature of all such operators is that they have, at minimum, a $|\mathcal{H}^{(B)}|$-fold degeneracy of all eigenvalues because of this $\mathbb{I}^{(B)}$.
It is also true that for any operator, given only its spectrum, one can infer the size of the smallest possible tensor factor on which it can be considered to act locally under \textit{any} tensor product factorization of the Hilbert space.  We refer to this property as localizability, and we denote the dimension of this smallest possible tensor factor for some operator $\hat{\mathcal{O}}$ as $\mathcal{D}_{\hat{\mathcal{O}}}$. This minumum size, $\mathcal{D}_{\hat{\mathcal{O}}}$, is simply determined as the total Hilbert space dimension $|\mathcal{H}|$ divided by the greatest common divisor of the multiplicity of the operator eigenvalues:
\begin{equation}
    \mathcal{D}_{\hat{\mathcal{O}}}:=|\mathcal{H}|/\text{gcd}(\{d_{\mathcal{O}_\lambda}\}),
\end{equation}
where $\mathcal{O}_\lambda$ are the eigenvalues of $\hat{\mathcal{O}}$ and $d_{\mathcal{O}_\lambda}$ are the multiplicities of those eigenvalues.  Of course, for a generic operator with nondegenerate spectrum, $\mathcal{D}_{\hat{\mathcal{O}}}$ will be the full dimension of the Hilbert space $|\mathcal{H}|$, and it cannot be localized onto any subfactor.  But in many contexts, operators of concern are local or localizeable in this manner.\footnote{It is may also be useful to consider operators which are ``approximately'' localizeable, but we consider only the exact case in this work.  We will also discuss some other means of lifting this restriction in the discussion and conclusions.}   We refer to any basis in which the operator $\hat{\mathcal{O}}$ takes the form $\eqref{eq: local op}$ where $|\mathcal{H}^{(A)}|$ has been made as small as possible, as a basis in which $\hat{\mathcal{O}}$ has been localized to $A$ (equivalently, $\mathcal{H}^{(A)}$).
\footnote{The use of spectral properties to determine locality structure is somewhat inspired by \cite{Cotler:2017abq}, but the pertinent notion of locality in that work was ``$k-$locality.''  An operator is $k-$local if it can be written as a sum of terms each acting on $\le k$ factors in some tensor product structure on the Hilbert space. A ``tensor product structure'' connotes a partitioning of the Hilbert space into an arbitrary number of small tensor factors, whereas in this work we only emphasize various Hilbert space bipartitions into subsystems $A$ and $B$.   An operator that is not $\mathcal{D}_{\hat{\mathcal{O}}}$-localizeable on any factor smaller than the full Hilbert space may still be $k-$local for some small $k$.  While $\mathcal{D}_{\hat{\mathcal{O}}}$-localizeability may be immediately ruled out if the operator lacks the appropriate degeneracy structure, the question of whether it is still $k-$local for some small $k$ is a much harder question, which we do not comment on here.}

One way to think about localizeability is as follows.  Given an operator $\hat{\mathcal{O}}$ specified in any bases, we may consider all possible unitaries $U$ on $\mathcal{H}$ as enacting a change of basis on the Hilbert space.  If the operator is localizeable, some subset of these bases will put the operator into the form \eqref{eq: local op} under some bipartition of the Hilbert space into subsystems $A$ and $B$. Some further subset of these will make the subsystem $A$ as small as possible while consistent with the operator spectrum.  We call any basis that accomplishes this a `localizing' basis for $\mathcal{O}$.  Such a basis is not unique.  Unitaries of the form $U^{(A)}\otimes U^{(B)}$ preserve a tensor product factorization of the form \eqref{eq: bipartite hilbert space}. Thought of as a change of basis, these take one localizing basis to another.  There are also nonlocal unitaries (nonlocal across this bipartition) that control on the eigensectors of $\hat{\mathcal{O}}^{(A)}$, enacting a different unitary on $\mathcal{H}^{(B)}$ for each eigensector of $\hat{\mathcal{O}}^{(A)}$.  As a change of basis, these \textit{do} alter the Hilbert space bipartition, but leave the operator $O$ still localized in the resulting basis.\footnote{This can be thought of as changing the way that the sole operator $\hat{\mathcal{O}}$ is incorporated into a type I factor algebra, the set of bounded operators on $\mathcal{H}^{(A)}$.}  These non-uniquenesses are not important for our purposes.  It suffices to identify any localizing basis for $\mathcal{O}$ and proceed as in the next section.

If one already knows the full structure of the operator, including not only its spectrum but also its eigenstates in some basis, then constructing a localizing basis is simply a matter of sorting this data.  In most circumstances, a Hamiltonian system is specified with an in-built tensor product structure and an associated `laboratory basis' in which the operators of interest are local with respect to the corresponding subsystems.  Note, however, that some operators can be localized onto much smaller tensor factors than their locality structure in the laboratory basis would suggest.  For instance, a product of five Pauli operators on different sites of a spin chain may certainly be considered local on those five sites.  But based on its spectrum alone, it may also be considered local to a much smaller factor, in fact a single qubit factor, in some other tensor product factorization of the Hilbert space.  Making $\mathcal{H}^{(A)}$ as small as possible serves to isolate the parts of the Hamiltonian in a decomposition that the operator directly `cares about' if it is to be regarded as part of a small, open subsystem. This allows us to concretely assess the intuition that operators on a subsystem thermalize whenever the complement system serves as a bath, and the interactions with that bath are suitable.  Our goal in the next question is to quantitatively relate such intuition to the ETH ansatz, and particularly its implications for the off-diagonal behavior of matrix elements.

\subsection{An operator-specific perturbation problem}
\label{subsec: operator-specific perturbation}
Given an operator $\hat{\mathcal{O}}=\hat{\mathcal{O}}^{(A)}\otimes \mathbb{I}^{(B)}$ that is localized to a subsystem $A$ as described above, an obvious decomposition of the Hamiltonian suggests itself as particularly relevant to the thermalization of that operator.  We may divide the Hamiltonian into pieces which act on factors $\mathcal{H}^{(A)}$ and $\mathcal{H}^{(B)}$ separately, and a term that interacts these two factors.  We treat the latter piece as an interaction term $\hat{H}^{(I)}$ that perturbs a decoupled Hamiltonian $\hat{H}^{(0)}$:
\begin{equation}\begin{split}\label{eq: hamiltonian decomposition}
	\hat{H}^{(T)} &:= \hat{H}^{(0)}+ \hat{H}^{(I)},\\
	\hat{H}^{(0)}&:=\hat{H}^{(A)}\otimes \hat{\mathbb{I}}^{(B)} +\hat{\mathbb{I}}^{(A)}\otimes  \hat{H}^{(B)}
    =\hat{H}^{(A)} +\hat{H}^{(B)},
\end{split}\end{equation}
where the notation after the last equality leaves identity factors implicit.  Superscripts on these Hamiltonians specify the total, bare, and interaction Hamiltonians as $\hat{H}^{(T)}$, $\hat{H}^{(0)}$, and $\hat{H}^{(I)}$, respectively, while $\hat{H}^{(A)}$ and $\hat{H}^{(B)}$ denote the decoupled hamiltonians that act separately on factors $\mathcal{H}^{(A)}$ and $\mathcal{H}^{(B)}$, and which together comprise the bare Hamiltonian $\hat{H}^{(0)}$.  

We emphasize that the Hamiltonian decomposition \eqref{eq: hamiltonian decomposition} is `induced' by the operator $\hat{\mathcal{O}}$ under consideration, while other operators suggest different decompositions.  The same decomposition is relevant to the full algebra of operators on the same factor $\mathcal{H}^{(A)}$, although other elements of that algebra may be localizeable onto yet smaller factors, whereas $\mathcal{H}^{(A)}$ has been made as small as possible for $\hat{\mathcal{O}}$.  For systems with a translation invariance, the decomposition \eqref{eq: hamiltonian decomposition} may be effectively equivalent for operators on a whole family of similar subfactors, but this need not be the case in general. Note that there is also no reason, in general, that the interaction Hamiltonian resulting from such a decomposition will be small (e.g.\ in operator norm) compared to the decoupled Hamiltonian(s). For instance, if the total Hamiltonian were randomly selected from a GUE distribution, under a typical bipartition of the Hilbert space the interaction Hamiltonian $\hat{H}^{(I)}$ will be parametrically larger than $\hat{H}^{(0)}$.  It is useful to restrict attention first to cases where the interaction term is small, and so it may fairly be called a ``perturbation'' to $\hat{H}^{(0)}$.  In particular, we require that it does not substantially alter the density of states from $\hat{H}^{(0)}$.  We will make a few comments on the more general case again below.

We now proceed as in Deutsch's analysis of section \ref{subsec: Deutschs approach}, leveraging the fact that the unperturbed Hamiltonian $\hat{H}^{(0)}$ is decoupled between $A$ and $B$ factors, so its eigenstates are simple products of $\hat{H}^{(A)}$ and $\hat{H}^{(B)}$ eigenstates:
\begin{equation}\begin{split}
		&\hat{H}^{(0)}\ket*{E_i^{(A)}, E_j^{(B)}} = (E^{(A)}_i + E_j^{(B)})\ket*{E_i^{(A)}, E_j^{(B)}},\\
		&\hat{H}^{(A)}\ket*{E_i^{(A)}} = E^{(A)}_i\ket*{E_i^{(A)}} , \hspace{4mm}\hat{H}^{(B)}\ket*{E_j^{(B)}} = E^{(B)}_j\ket*{E_j^{(B)}}.
\end{split}\end{equation}
We use the notation that $\ket*{E_i^{(A)}, E_j^{(B)}} :=\ket*{E_i^{(A)}}\otimes\ket*{E_j^{(B)}}$ for positive integers $i\le |\mathcal{H}^{(A)}|$ and $j\le |\mathcal{H}^{(B)}|$.  We denote the eigenstates of the total Hamiltonian as $\ket*{E^{(T)}_\alpha}$, where $	\hat{H}^{(T)}\ket*{E_\alpha^{(T)}} = E^{(T)}_\alpha \ket*{E_\alpha^{(T)}}$ for positive integer $\alpha\le|\mathcal{H}^{(T)}|$, since these generically have no bipartite structure.  We can then express each of these total eigenstates in the basis of the unperturbed Hamiltonian eigenstates as
\begin{equation}\begin{split}\label{eq: scrambled coefficients}
	\ket*{E_\alpha^{(T)}} = \sum_{i}^{|\mathcal{H}^{(A)}|}\sum^{|\mathcal{H}^{(B)}|}_{j}c^{\alpha}_{ij}\ket*{E^{(A)}_i, E^{(B)}_j}.
\end{split}\end{equation}
The supercript $\alpha$ on $c^{\alpha}_{ij}$ labels which total eigenstate is being expressed, while the indices $i$ and $j$ run over the non-interacting Hamiltonian eigenstates of the two factors.  Here and in what follows, we use Greek letters for indices on the total Hilbert space and Latin indices on factors of the bipartition.  There is no fundamental difference between raised and lowered indices; we use raised indices when we find it convenient for clarity.

We will refer to the coefficients $c^{\alpha}_{ij}$ as the `scrambling coefficients', because the general expectation is that in a chaotic regime the total system eigenstates will consist of a `scrambled' superposition of the unperturbed eigenstates around a similar energy.  This general phenomenon is illustrated in figure \ref{fig: coefficient scrambling ising}, for a system introduced in that section.  Finally, we express the matrix elements of the operator $\hat{\mathcal{O}}$ in terms of the basis of total energy eigenstates:
\begin{equation}\label{eq: exact form}
	\bra*{E^{(T)}_\alpha}\hat{\mathcal{O}}\ket*{E^{(T)}_\beta} = 
	\sum_{i,j}^{|\mathcal{H}^{(A)}|}\sum^{|\mathcal{H}^{(B)}|}_{k}c^{\star\alpha}_{ik}c^{\beta}_{jk}\bra*{E^{(A)}_i}\hat{\mathcal{O}}^{(A)} \ket*{E^{(A)}_j}.
\end{equation}
This is directly analogous to equation \eqref{eq: exact Deutsch} in Deutsch's method.  The local nature of the operator now makes the sum over the $B$ factor trivial.  At this point, in any real system the details of the relevant perturbation problem \eqref{eq: hamiltonian decomposition} for operators of interest may be studied to arrive at a detailed ansatz for the coefficients $c^{\alpha}_{ij}$.  In the next section we will explore the implications of a few simple ansatzes. 

\section{Scrambling ansatzes}
\label{sec: scrambling ansatzes}
A general expectation for perturbation problems of the sort described in the preceding section is that, in chaotic regimes, the eigenstates of $\hat{H}^{(T)}$ will consist of a scrambled superposition of eigenstates of $\hat{H}^{(0)}$ at similar energy.  This `approximate microcanonical scrambling' is illustrated in figure \ref{fig: coefficient scrambling ising} for a chaotic spin chain described in that section.  Specifying the statistics the scrambling coefficients $c^{\alpha}_{ij}$ in equation \eqref{eq: scrambled coefficients} for a class of perturbations in a given system allows us to study the onset of ETH and its deviations in that system.  Here we will explore the implications of a few simple scrambling ansatzes.  Essentially we ask, when some form of idealized local eigenstate scrambling does occur, what does the locality of the operator imply about the ETH structure of its matrix elements?

\subsection{Microcanonical scrambling}
\label{subsec: microcanonical ansatz}
First, suppose that there is ``perfect scrambling'' across microcanonical windows of characteristic width $\Delta\sim ||\hat{H}^{(I)}||$.  By this we mean that the coefficients $c^{\alpha}_{ij}$ express a Haar random vector in the sub-Hilbert space built from unperturbed eigenstates in an energy window around $E^{(T)}_\alpha$.  We denote this subspace as $\mathcal{H}^{(0)}_{E^{(T)}_\alpha}$.  The coefficients $c^{\alpha}_{ij}$ are then treated as independent gaussian variables of mean zero and variance $|\mathcal{H}^{(0)}_{E^{(T)}_\alpha}|^{-1}$ when $|E^{(T)}_\alpha-E^{(A)}_i-E^{(B)}_j|\le \Delta/2$, and vanishing outside this range:  
\begin{equation}\begin{split}
    \overline{c^{\alpha}_{ij}}:=0,\qquad\quad
    \overline{|c^{\alpha}_{ij}|^2}:=
    \begin{cases}
        \big|\mathcal{H}^{(0)}_{E^{(T)}_{\alpha}}\big|^{-1}
        &|E^{(T)}_\alpha-E^{(A)}_i-E^{(B)}_j|\le \Delta/2\\
        0&\text{ otherwise}
    \end{cases},
\end{split}\end{equation}
The overbar may be thought of as an ensemble average over similar perturbations.
Such a model is too crude to capture any physics associated with the tails of any realistic distribution (because it explicitly eliminates such tails), but it provides some clean intuition which persists under more realistic ansatzes in a limit of narrow scrambling.  Assuming such a distribution leads us to replace the sum of coefficients appearing in \eqref{eq: exact form} with the ansatz\footnote{In this replacement and similar ones below, we essentially aim to recover only the expected means and variances appropriate to any ansatz, remaining agnostic about higher moments, which in principle alter the more detailed statistics of the $R^{\alpha\beta}_{ij}$.}
\begin{equation}\begin{split}\label{eq: microcanonical scrambling ansatz}
	\sum_k^{|\mathcal{H}^{(B)}|}c_{ik}^{*\alpha}c^{\beta}_{jk} \approx
	\delta^{\alpha\beta}
	\delta_{ij}\frac{\big|\mathcal{H}^{(B)}_{E_\alpha^{(T)}-E_i^{(A)}}\big|}{\big|\mathcal{H}^{(0)}_{E_\alpha^{(T)}}\big|}
	+R^{\alpha\beta}_{i j}\sqrt{\frac{\big| \mathcal{H}^{(B)}_{E_\alpha^{(T)}-E_i^{(A)}} \cap \mathcal{H}^{(B)}_{E^{(T)}_\beta-E^{(A)}_j}\big|}{\big|\mathcal{H}^{(0)}_{E^{(T)}_\alpha}\big|\big|\mathcal{H}^{(0)}_{E^{(T)}_\beta}\big|}}.
\end{split}\end{equation}
Here, each $\mathcal{H}^{(B,0)}_E$ (for either $(B)$ or $(0)$ superscript, and for any subscript value $E$) denotes the microcanonical sub-Hilbert space of $\mathcal{H}^{(B,0)}$ spanned by eigenstates of $\hat{H}^{(B,0)}$ in the energy range $[E-\Delta/2,E+\Delta/2]$. The symbol $\cap$ denotes the intersection of two such microcanonical windows, and the corresponding sub-Hilbert space.  Once again, there is no fundamental difference between raised and lowered indices (e.g.\, $\delta^{\alpha\beta}=\delta_{\alpha\beta}$ is a Kronecker delta function) except to recall the associations of these indices with coefficients on the left hand side of \eqref{eq: microcanonical scrambling ansatz}. One can check that the coefficient ansatz on the right-hand side of \eqref{eq: microcanonical scrambling ansatz} replicates the means and variances of the sum on the left hand side under ``microcanonical scrambling'' assumptions if $R^{\alpha\beta}_{i j}$ is a random (generalized) matrix with complex elements of mean zero and variance one, satisfying $R^{\alpha\beta}_{i j} = R^{\beta\alpha\star}_{j i}$.\footnote{Analogous statements for the case of real coefficients $c^{\alpha}_{ij}$, as appropriate for a system with time-reversal invariance, are again obtainable through minor adjustments.}

\subsubsection{Microcanonical scrambling: diagonal contribution}
\label{subsec: microcanonical ansatz - diagonal}
Focusing first on the implications of this ansatz for the diagonal matrix elements, note that the ratio of sub-Hilbert spaces appearing next to the $\delta^{\alpha\beta}\delta_{ij}$ in \eqref{eq: microcanonical scrambling ansatz} is the same ratio that commonly appears in derivations of subsystem Gibbs states from global microcanonical state, which are based on counting the number of microstates compatible with a fixed, small window of total energy under exchanges of energy between a subsystem and environment (here $A$ and $B$): 
\begin{equation}
	\delta^{\alpha\beta}\delta_{ij}\frac{\big|\mathcal{H}^{(B)}_{E_\alpha^{(0)}-E_i^{(A)}}\big|}{\big|\mathcal{H}^{(0)}_{E_\alpha^{(T)}}\big|}
	\approx 
	\delta^{\alpha\beta}\delta_{ij}\frac{\exp(-\beta_\alpha E^{(A)}_i)}{\sum_l \exp(-\beta_\alpha E^{(A)}_l)}
	=\delta^{\alpha\beta}\delta_{ij}\bra*{E^{(A)}_i}\hat{\rho}_{\text{Gibbs}}^{(A)}(\beta_\alpha)\ket*{E^{(A)}_i},
\end{equation}
where $\hat{\rho}_{\text{Gibbs}}^{(A)}(\beta):=\exp(-\beta\hat{H}^{(A)})/\tr(\exp(-\beta\hat{H}^{(A)}))$, and  $\beta_\alpha:=\beta(E^{(T)}_\alpha)$ is the inverse temperature set by requiring $\bra*{E^{(T)}_\alpha}\hat{H}^{(A)}\otimes\mathbb{I}^{(B)}\ket*{E^{(T)}_\alpha}=\tr^{(A)}\left[ \hat{\rho}_{\text{Gibbs}}^{(A)}\left(\beta_\alpha\right)\hat{H}^{(A)}\right]$.   This behavior is essentially built-in to the ansatz \eqref{eq: microcanonical scrambling ansatz} as a manifestation of ``canonical typicality'' \cite{Goldstein_2006}, which is the statement that in quantum systems under some weak interaction assumptions, not only do microcanonical ensembles of the full system reduce to a Gibbs states on small subsystems, but so do most/typical \textit{pure} states drawn from a microcanonical window (we refer the reader to \cite{Goldstein_2006} for more details).  The result is that applying the ansatz \eqref{eq: microcanonical scrambling ansatz} in \eqref{eq: exact form}, when the operator is local on a small factor $|\mathcal{H}^{(A)}|<<|\mathcal{H}^{(B)}|$, the first (diagonal) term directly contributes a Gibbs-state expectation value of the operator $\hat{\mathcal{O}}^{(A)}$.  

\begin{equation}\begin{split}\label{eq: microcanonical scrambling diag result}
	 \sum_{ij}^{|\mathcal{H}^{(A)}|}\delta_{\alpha\beta}\delta_{ij}\frac{\big|\mathcal{H}^{(B)}_{E_\alpha^{(0)}-E_i^{(A)}}\big|}{\big|\mathcal{H}^{(0)}_{E_\alpha^{(T)}}\big|}
     \hat{\mathcal{O}}^{(A)}_{ij}
     &\approx 
	\delta_{\alpha\beta}\tr\left(\hat{\rho}_{\text{Gibbs}}^{(A)}(\beta_\alpha)\hat{\mathcal{O}}^{(A)}\right).\\
\end{split}\end{equation}

\subsubsection{Microcanonical scrambling: off-diagonal ansatz}
\label{subsec: microcanonical ansatz - off diagonal}
Considering now the implications of the ansatz \eqref{eq: microcanonical scrambling ansatz} for off-diagonal matrix elements $(\alpha\ne\beta)$, which leads to
\begin{equation}\begin{split}\label{eq: microcanonical scrambling off-diag result}
	\bra*{E^{(T)}_\alpha}\hat{\mathcal{O}}\ket*{E^{(T)}_\beta} &\approx 
	R_{\alpha\beta}\sqrt{\sum_{i,j}^{|\mathcal{H}^{(A)}|}\frac{\big|\mathcal{H}^{(B)}_{E^{(T)}_\alpha-E^{(A)}_i}\cap \mathcal{H}^{(B)}_{E^{(T)}_\beta-E^{(A)}_j}\big|}{|\mathcal{H}^{(0)}_{E_\alpha}||\mathcal{H}^{(0)}_{E_\beta}|}|\hat{O}^{(A)}_{ij}|^2}.
\end{split}\end{equation}
The $R_{\alpha\beta}$ is again a random matrix with elements of mean zero and variance one, satisfying $R_{\alpha\beta}=R^\star_{\beta\alpha}$. 
The precise definition of the microcanonical subspaces in \eqref{eq: microcanonical scrambling off-diag result} depends on the width of the scrambling window, which we take to be set by the interaction Hamiltonian $\Delta\sim ||\hat{H}^{(I)}||$.
We make a few conceptual observations at this point. First, note that if the scrambling window is taken to be arbitrarily large (e.g. if the $||\hat{H}^{(I)}||$ is on the same order or larger than $||\hat{H}^{(0)}||$, violating our weak interaction assumptions), the random matrix result \eqref{eq: RMT result} is recovered, because the quantity under the radical sign becomes $|\mathcal{H}^{(B)}||\mathcal{H}^{(0)}|^{-2}\sum_{ij}|\hat{\mathcal{O}}^{(A)}_{ij}|^2 = \overline{\mathcal{O}^2}/|\mathcal{H}^{(0)}|$.  By contrast, for finite $\Delta$ the matrix elements vanish \textit{identically} when $|E^{(T)}_{\alpha}-E^{(T)}_\beta|  > \Delta+\sigma_A$, where $\sigma_A$ is the spectral range of the $A$ Hamiltonian.  This is because the overlaps $\big| \mathcal{H}^{(B)}_{E_\alpha^{(T)}-E_i^{(A)}} \cap \mathcal{H}^{(B)}_{E^{(T)}_\beta-E^{(A)}_j}\big|$ always vanish identically for such $\omega^{(T)}_{\alpha\beta}$.  This is an artifact of the ``perfect'' microcanonical scrambling ansatz, which eliminates any tails around the scrambling window, but even with more realistic ansatzes we expect strong suppression for $|\omega^{(T)}_{\alpha\beta}|\gtrsim \frac{1}{2}(\Delta + \sigma_A)$. (Recall again that $\Delta$ is the characteristic width of the scrambling and $\sigma_A$ is the spectral range of $\hat{H}^{(A)}$.)  Any predictions of the `perfect microcanonical scrambling' assumption can only be relevant in a narrow scrambling regime, where the detailed shape of the scrambling window becomes irrelevant.    

Another interesting observation is that in the regime $\omega_{\alpha\beta} \lesssim \sigma_A$ the overlap $\big| \mathcal{H}^{(B)}_{E_\alpha^{(T)}-E_i^{(A)}} \cap \mathcal{H}^{(B)}_{E^{(T)}_\beta-E^{(A)}_j}\big|$ peaks when $E^{(T)}_{\alpha}-E^{(T)}_\beta =E^{(A)}_i-E^{(A)}_j$.  In a regime where the scrambling width $\Delta$ is small even compared to the spectral gaps of $\hat{H}^{(A)}$, this peaking gives a special role to these spectral gaps: they serve as a `filter' against which the matrix elements $\hat{O}^{(A)}_{ij}$ are summed, imprinting a banding structure on near off-diagonal matrix elements.  For small systems and narrow scrambling window this affect can be quite pronounced, as we show in appendix \ref{app: random hamiltonian systems} using a random matrix model tuned for that purpose.  It is worth emphasizing that the spectral gaps referred to here are those of $\hat{H}^{(A)}$, so when $A$ is a small subsystem these are not exponentially small in the \textit{total} system size.  Even so, the regime referred to here is either the very near-diagonal regime, or a system/regime where the Hamiltonian $\hat{H}^{(A)}$ has very different properties from the $\hat{H}^{(B)}$ Hamiltonian. For now we return to discussion of more standard ETH-like regimes. 

As soon as the subsystems $A$ and $B$ are even as large as a few qubits, it is useful to move to a continuum approximation of the spectra.  We now assume that the density of eigenstates of $\hat{H}^{(C)}$ for $C\in \{A,B,0,T\}$ can be represented by smooth spectral density functions $n_C(\epsilon)$ such that $n_{C}(\epsilon)\dd \epsilon$ represents the number of eigenstates in the range $[\epsilon,\epsilon+\dd\epsilon]$, and $\int_{-\infty}^\infty \dd \epsilon n_{C}(\epsilon) = |\mathcal{H}^{(C)}|$.  We sometimes alternatively employ a fractional density of states $\rho_C(\epsilon)=|\mathcal{H}^{(C)}|^{-1}n_C(\epsilon)$ such that $\int \dd\epsilon \rho_C(\epsilon)=1$.
\begin{equation}
    \sum_i^{\hat{H}^{(C)}}\rightarrow \int \dd\epsilon~ n_C(\epsilon) = |\mathcal{H}^{(C)}|\int \dd\epsilon~\rho_c(\epsilon), \qquad C\in\{A, B, 0, T\}. 
\end{equation}
Integrations are always taken over the full real line with the understanding that these spectral density functions vanish outside of their respective finite spectral ranges $\sigma_C$.\footnote{In a slight abuse of notation, we sometimes use $\sigma_C$ to denote the domain of nonzero $\rho_C$, i.e. $\sigma_C:=[ \epsilon_C^{\text{min}},\epsilon_C^{\text{max}}]$, but usually we mean the width of this interval, $\sigma_C:=\epsilon_C^{\text{max}}-\epsilon_C^{\text{min}}$.  The meaning should be clear from context.} If the $A$-subsystem is taken to be a very small, discrete sums on $\mathcal{H}^{(A)}$ may be more appropriate; we will convert between such expressions as needed.  

Three relevant scales to consider are now the spectral ranges $\sigma_A$ and $\sigma_B$ (alternatively, $\sigma_0=\sigma_A+\sigma_B$) parametrizing the relative sizes of $A$ and $B$ factors, and the scrambling width $\Delta$, which we will usually trade for the corresponding variance $\sigma_S = \Delta/(2\sqrt{3})$ of the flat scrambling distribution.  If the density of states functions are relatively featureless we may consider these the \textit{only} relevant scales.  Regardless, the perfect microcanonical scrambling ansatz should provide a robust prediction when the scrambling width $\Delta\sim \sigma_s\sim ||\hat{H}^{(I)}||$ is small compared to the scales on which $n_A$ and $n_B$ vary. This might be considered a prototypical ETH regime for systems with local interactions, where scaling up the size of $|\mathcal{H}^{(A)}|$ and $|\mathcal{H}^{(B)}|$ naturally sends the scrambling variance $\sigma_S\sim ||\hat{H}^{(I)}||$ much smaller than $\sigma_A$ and $\sigma_B$. 

To further use \eqref{eq: microcanonical scrambling off-diag result} to understand the behavior of off-diagonal elements $\bra*{E^{(T)}_\alpha}\hat{\mathcal{O}}^{(A)}\ket*{E^{(T)}_\beta}$, which should specify what sort of local operators $\hat{\mathcal{O}}^{(A)}$ we are interested in.  By construction, these operators are already as local as possible (meaning they cannot be localized on any further subfactor), so we cannot use any locality considerations to argue that they should be of ETH form on the subfactor $A$.  It may be useful to consider various forms of structured matrix elements $\mathcal{O}^{(A)}_{ij}$ in specific contexts.  Appendix \ref{app: approximations from perfect microcanonical scrambling} shows some more general expressions, but here we report generic behavior of `typical' operators that are local to the $A$ factor (see equation \eqref{eq: RMT result}), and so we replace $|\mathcal{O}^{(A)}_{ij}|^2\rightarrow \overline{\mathcal{O}^{2(A)}}/|\mathcal{H}^{(A)}|$, where $\overline{\mathcal{O}^{2(A)}}=\overline{\mathcal{O}^{2}}$ represents the average of the squared spectrum of $\hat{\mathcal{O}}^{(A)}$.  We show in appendix \ref{app: approximations from perfect microcanonical scrambling} that the leading behavior in a narrow scrambling regime is
\begin{equation}\begin{split}\label{eq: narrow scrambling limit}
    \bra*{E^{(T)}_\alpha}\hat{\mathcal{O}}\ket*{E^{(T)}_\beta} &\approx R_{\alpha\beta}\sqrt{\frac{\overline{\mathcal{O}^2}}{|\mathcal{H}^{(A)}|}\frac{\int \dd\epsilon ~n_A(\epsilon+\omega^{(T)}_{\alpha\beta})n_A(\epsilon-\omega^{(T)}_{\alpha\beta})
    n_B(\bar{E}^{(T)}_{\alpha\beta}-\epsilon)}{n_0(E^{(T)}_\alpha)n_0(E^{(T)}_\beta)}},
    \qquad \sigma_S<<\sigma_A,\sigma_B.
\end{split}\end{equation}
To relate this expression back to the standard ETH ansatz \eqref{eq: ETH ansatz}, we can define an entropic suppression factor as $e^{-S(\bar{E}^{(T)}_{\alpha\beta})/2}:=(\sigma_S n_0(\bar{E}^{(T)}_{\alpha\beta}))^{-1/2}$, and then write
\begin{equation}\begin{split}\label{eq: f narrow scrambling limit}
    f_{\hat{\mathcal{O}}}(\bar{E},\omega) &\approx \sqrt{\frac{\overline{\mathcal{O}^2}}{|\mathcal{H}^{(A)}|}\frac{\sigma_S n_0(\bar{E})}{n_0(\bar{E}+\omega)n_0(\bar{E}-\omega)}\int \dd\epsilon ~n_A(\epsilon+\omega)n_A(\epsilon-\omega)
    n_B(\bar{E}-\epsilon)},
    \qquad \sigma_S<<\sigma_A,\sigma_B,
\end{split}\end{equation}
Of course, an overall constant is sensitive to the precise  definition of entropic suppression.

If we further restrict to a regime where the $A$-factor is a small subsystem, such that $\sigma_A<<\sigma_B$, and if the density of states $n_B$ is approximately fixed in the interval $\bar{E}\pm\sigma_A/2$, then \eqref{eq: f narrow scrambling limit} simplifies to
\begin{equation}\begin{split}\label{eq: f narrow scrambling small A limit}
    f_{\hat{\mathcal{O}}}(\bar{E},\omega) 
    &\approx \sqrt{\overline{\mathcal{O}^2} \sigma_S ~[\rho_A\star\rho_A](2\omega)},
    \qquad \sigma_S<<\sigma_A<<\sigma_B,
\end{split}\end{equation}
where $[g_1\star g_2](x):=\int \dd y g_1(y)g_2(y+x)$ denotes a cross-correlation of two real functions $g_1$ and $g_2$.
Here we've replaced $n_A$ with the fractional density of states $\rho_A:=|\mathcal{H}^{(A)}|^{-1}n_A$.  We see that the off-diagonal dependence of typical operators on the $A$ factor in this regime is simply related to an autocorrelation of the $\hat{H}^{(A)}$ density of states, and $f_\mathcal{O}$ identically vanishes for $|\omega| > \sigma_A/2$.  Note that in this limit the expression has become independent of $\bar{E}$, so that the only $\bar{E}$-dependence of the off-diagonal matrix elements is restricted to the entropic suppression factor which we peeled off after equation \eqref{eq: narrow scrambling limit}.  
For any relatively featureless density of states $\rho_A$, a reasonable first approximation is to consider a flat distribution over the spectral range $\sigma_A$, in which case we have $[\rho_A\star \rho_A](2\omega) = \frac{1}{\sigma_A}\left(1-\frac{2|\omega|}{\sigma_A}\right)\Theta(\sigma_A/2-|\omega|)$, with $\Theta$ being a Heaviside step function, and \eqref{eq: f narrow scrambling small A limit} becomes
\begin{equation}\begin{split}\label{eq: f narrow scrambling small A limit, flat A}
    f_{\hat{\mathcal{O}}}(\bar{E},\omega) 
    &\approx \Theta\left(1-\frac{2|\omega|}{\sigma_A}\right)\sqrt{\overline{\mathcal{O}^2}
    \frac{\sigma_S}{\sigma_A}\left(1-\frac{2|\omega|}{\sigma_A}\right)},
    \qquad \sigma_S<<\sigma_A<<\sigma_B,
\end{split}\end{equation}
resulting in a simple linear decay of $|f_{\hat{\mathcal{O}}}(\omega)|^2$ as $|\omega|$ increases up until $\sigma_A/2$, after which point $f_{\hat{\mathcal{O}}}$ vanishes identically.

To recap, equations \eqref{eq: f narrow scrambling limit}, \eqref{eq: f narrow scrambling small A limit}, and \eqref{eq: f narrow scrambling small A limit, flat A} are successive approximations for `typical' operators on the $A$-factor in the limit of narrow scrambling $(\sigma_S<<\sigma_A, \sigma_B)$, small subsystems $A$ $(\sigma_S<<\sigma_A<<\sigma_B)$, and a featureless (flat) spectral density of the $\hat{H}^{(A)}$ Hamiltonian.  We have not subjected such results to a wide variety of numerical tests in a varied systems, but in section \ref{sec: numerical illustrations} we find good agreement in the case of a 1-dimensional chaotic spin chain. 

If we wish to move away from the ``narrow'' scrambling limit, as may be required by the relevant perturbation problem, it is useful to characterize the affects of a more realistic scrambling ansatzes, such as those allowing tails around the scrambling energy as opposed to a perfectly sharp microcanonical distribution.  We turn to this next.

\subsection{Smooth scrambling}
\label{subsec: smooth ansatz}
Instead of modelling the coefficients $c^{\alpha}_{ij}$ as `perfectly' scrambled on a microcanonical subspace, we now consider a distribution such that the variance is peaked, but smoothly decaying, around the energy $E^{(T)}_\alpha$.  In other words, we continue to treat the coefficients as independent random variables, but now with a variance that is a function of the energy distance $|E^{(T)}_\alpha - E^{(A)}_i-E^{(B)}_j|$.\footnote{To fully leverage the bipartite nature of the perturbation problem, it would be interesting to consider scrambling ansatzes that treat the factors $\mathcal{H}^{(A)}$ and $\mathcal{H}^{B}$ less symmetrically, but we consider the simplest case for now.} With an overline denoting an ensemble average, we take
\begin{equation}\begin{split}
\label{eq: smooth coefficient ansatz}
    \overline{c^{\alpha}_{ij}}:=0,\qquad \overline{|c^{\alpha}_{ij}|^2} &=h(E_{\alpha}^{(T)}-E_i^{(A)}-E_j^{(B)})/Z(E^{(T)}_\alpha),
\end{split}\end{equation}
where $h(E)$ is some smooth, dimensionless function, peaked at $E=0$ and monotonically decaying to zero as $|E|\rightarrow\pm\infty$ (for example, when considering the specifics of the perturbation problem it might be found that a Gaussian of width $\sim||\hat{H}^I||$ is appropriate, but we leave $h$ general for now).  The function $Z(E)$ is as required by normalization:
\begin{equation}\begin{split}
\label{eq: discrete Z}
    Z(E)&:=
    \sum_i^{|\mathcal{H}^{(A)}|}\sum_j^{|\mathcal{H}^{(B)}|}
    h(E-E_i^{(A)}-E_j^{(B)}).\\
\end{split}\end{equation}
This motivates the replacement
\begin{equation}\begin{split}
\label{eq: discrete ansatz}
    \sum_k^{|\mathcal{H}^{(B)}|}c^{\alpha*}_{ik}c^{\beta}_{jk}
    &\rightarrow \sum_k^{|\mathcal{H}^{(B)}|}\frac{h(E_{\alpha}^{(T)}-E_i^{(A)}-E_k^{(B)})}{Z(E_\alpha^{(T)})}\delta_{ij}\delta^{\alpha\beta}
    +R^{\alpha\beta}_{ij}\sqrt{\frac{\sum_k^{|\mathcal{H}^{(B)}|}h(E_\alpha^{(T)}-E_i^{(A)}-E_k^{(B)})h(E_\beta^{(T)}-E_j^{(A)}-E_k^{(B)})}{Z(E_\alpha^{(T)})Z(E_\beta^{(T)})}},\\
\end{split}\end{equation}
where $R^{\alpha\beta}_{ij}$ is again a pseudorandom matrix with elements of mean zero and variance one, uncorrelated except that they satisfy $R^{*\alpha\beta }_{ij}=R^{\beta\alpha}_{ji}$.  Recall that there is no fundamental difference between raised and lowered indices except to recall the association of these indices with the coefficients on the left hand side of \eqref{eq: discrete ansatz}.  Employing this in \eqref{eq: exact form} then leads to the following ansatz for matrix elements:
\begin{equation}\begin{split}\label{eq: discrete result}
	\bra*{E^{(T)}_\alpha}\hat{\mathcal{O}}\ket*{E^{(T)}_\beta} \approx &
	\delta_{\alpha\beta}
    \sum_i^{|\mathcal{H}^{(A)}|}\sum_k^{|\mathcal{H}^{(B)}|}
    \frac{h(E^{(T)}_\alpha-E^{(A)}_i-E^{(B)}_k)}{Z(E^{(T)}_\alpha)} \hat{\mathcal{O}}^{(A)}_{ii}\\
	&+R_{\alpha\beta}
    \sqrt{\frac{\sum_{i,j}^{|\mathcal{H}^{(A)}|} \sum_k^{|\mathcal{H}^{(B)}|}h(E_\alpha^{(T)}-E_i^{(A)}-E_k^{(B)})h(E_\beta^{(T)}-E_j^{(A)}-E_k^{(B)})|\hat{\mathcal{O}}^{(A)}_{ij}|^2}{Z(E_\alpha^{(T)})Z(E_\beta^{(T)})}}.\\
\end{split}\end{equation}
where $R_{\alpha\beta}$ is again a random matrix with elements of mean zero and variance one, satisfying $R_{\alpha\beta}=R^\star_{\beta\alpha}$, and we've denoted the matrix elements of $\hat{\mathcal{O}}^{(A)}$ in the $\hat{H}^{(A)}$ eigenbasis as $\hat{O}^{(A)}_{ij}:=\bra*{E^{(A)}_i}\hat{O}^{(A)}\ket*{E^{(A)}_j}$.  We now consider separately the diagonal and off-diagonal contributions in \eqref{eq: discrete result}.

\subsubsection{Smooth scrambling: diagonal contribution}
\label{subsubsec: smooth ansatz - diagonal}
We first consider the first term in \eqref{eq: discrete result}, giving the smooth part of the diagonal contribution.  We again assume that the density of eigenstates of $\hat{H}^{(B)}$ can be represented by a smooth spectral density function $n_{B}(\epsilon)$ such that $n_{B}(\epsilon)\dd \epsilon$ represents the number of eigenstates in the range $[\epsilon,\epsilon+\dd\epsilon]$, and $\int_\infty^\infty \dd \epsilon~ n_{B}(\epsilon) = |\mathcal{H}^{(B)}|$.  We also now assume that the scale on which the function $h(E)$ is sharply peaked (at $E=0$) is narrower than the scale on which $n_B(\epsilon)$ varies.  Since this width of the peak is set by the interaction Hamiltonian, this is a weak interaction condition.   This allows us to express the sum over $k$ as
\begin{equation}
    \label{eq: h B sum}
    \sum_k^{|\mathcal{H}^{(B)}|}h(E-E^{(B)}_k)\approx
    \int \dd\epsilon n_B(\epsilon)h(E-\epsilon)\approx
    N_h n_B(E),\qquad N_h:=\int\dd\epsilon~ h(\epsilon).
\end{equation}
For sharply peaked and dimensionless $h(\epsilon)$, the constant $N_h$ expresses the characteristic energy range of the peaking. The smooth part of the diagonal contribution becomes
\begin{equation}\begin{split}\label{eq: discrete diagonal approx 1}
    \sum_i^{|\mathcal{H}^{(A)}|}\sum_k^{|\mathcal{H}^{(B)}|}
    \frac{h(E^{(T)}_\alpha-E^{(A)}_i-E^{(B)}_k)}{Z(E^{(T)}_\alpha)} \hat{\mathcal{O}}^{(A)}_{ii}\approx &
    \frac{\sum_i^{|\mathcal{H}^{(A)}|}n_B(E^{(T)}_\alpha-E^{(A)}_i)\hat{\mathcal{O}}^{(A)}_{ii}}{\sum_j^{|\mathcal{H}^{(A)}|}n_B(E^{(T)}_\alpha-E^{(A)}_j)}.\\
\end{split}\end{equation}
For small subsystems $A$, this again becomes a Gibbs expectation value.  Writing $s_B(E):= \ln(n_B(E))$, we have $\frac{n_B(E^{(T)}_\alpha-E^{(A)}_i)}{n_B(E^{(T)}_\alpha-E^{(A)}_j)} = e^{s_B(E^{(T)}_\alpha-E^{(A)}_i)-s_B(E^{(T)}_\alpha-E^{(A)}_i)}\approx e^{-(E^{(A)}_i-E^{(A)}_j)s'_B(E^{(T)}_A-\overline{E^{(A)}})}$ and the diagonal matrix elements become
\begin{equation}\begin{split}\label{eq: discrete diagonal Gibbs}
    \bra*{E^{(T)}_\alpha}\hat{\mathcal{O}}\ket*{E^{(T)}_\alpha} \approx &
    \frac{\sum_i e^{-\beta_\alpha E^{(A)}_i }\hat{\mathcal{O}}_{ii}}{\sum_j e^{-\beta_\alpha E^{(A)}_j}} + R_{\alpha\alpha}(...) \\
\end{split}\end{equation}
with $\beta_\alpha:=s'_B(E^{(T)}_\alpha-\overline{E^{(A)}})$.

Here our method of approximating the sum \eqref{eq: h B sum} using the sharply peaked nature of $h$ eliminates any dependence on the detailed form of $h$, so we haven't gone beyond the ``perfect microcanonical scrambling'' ansatz here. For now we'll be more interested in the off-diagonal contributions, which we turn to next.

\subsubsection{Smooth scrambling: off-diagonal and fluctuations}
\label{subsubsec: smooth ansatz - off diagonal}

For the off-diagonal contribution, we again start by assuming that the symmetric scrambling function $h$ is sharply peaked on the scale over which $n_B$ varies. For now we make no assumptions about the size of the $A$ subsystem.   This allows for the following approximations:
\begin{align}
    \label{eq: hh B sum}
    &\sum_k^{|\mathcal{H}^{(B)}|}h(E_1-E^{(B)}_k)h(E_2-E^{(B)}_k)\approx
    n_B((E_1+E_2)/2)[h\star h](E_1-E_2),
    \\
    \label{eq: Z approx}
    &Z(E)\approx
    N_h\int \dd\epsilon~ n_A(\epsilon)n_B(E-\epsilon)=
    N_h n_{0}(E),
\end{align}
where $[g_1\star g_2](x):= \int \dd y g_1(y)g_2(x+y)$ denotes the cross-correlation of two functions $g_1$ and $g_2$. Integrations can always be taken over the full real line with the understanding that the spectral densities vanish outside a their finite spectral ranges. Employing these approximations for the off-diagonal $\alpha\ne\beta$ part of \eqref{eq: discrete result} leads to
\begin{equation}\begin{split}\label{eq: smooth result off diagonal}
	&\bra*{E^{(T)}_\alpha}\hat{\mathcal{O}}\ket*{E^{(T)}_\beta} \approx 
	R_{\alpha\beta}\sqrt{\frac{\sum_{i,j}^{|\mathcal{H}^{(A)}|} n_B(\bar{E}^{(T)}_{\alpha\beta}-\bar{E}^{(A)}_{ij})[h\star h](2\omega^{(T)}_{\alpha\beta}-2\omega^{(A)}_{ij})|\hat{\mathcal{O}}^{(A)}_{ij}|^2}
    {N^2_h ~n_0(E_\alpha^{(T)})n_0(E_\beta^{(T)})}}.\\
\end{split}\end{equation}
Here we employ the notation
\begin{equation}\begin{split}\label{eq: change vars}
    \bar{E}^{(T)}_{\alpha\beta}:=\frac{E^{(T)}_\alpha+E^{(T)}_\beta}{2},\quad  \omega^{(T)}_{\alpha\beta}:=\frac{E^{(T)}_\alpha-E^{(T)}_\beta}{2},\\
    \bar{E}^{(A)}_{ij}:=\frac{E^{(A)}_i+E^{(A)}_j}{2},\quad  \omega^{(A)}_{ij}:=\frac{E^{(A)}_i-E^{(A)}_j}{2}.\\
\end{split}\end{equation}
Equation \eqref{eq: smooth result off diagonal} is the general prediction of a symmetric, sharply peaked scrambling ansatz $h$ (specifically, sharply peaked on the scale over which $n_B(\bar{E})$ varies).  Harking back to the standard ETH expression \eqref{eq: ETH ansatz}, we can define an overall entropic suppression factor of $e^{-S(\bar{E})/2}:=(\sigma_S n_0(\bar{E}))^{-1/2}$ (where $\sigma_S$ is the scrambling variance associated with $h$) and parametrize the remaining off-diagonal dependence in terms of a function $f_{\hat{\mathcal{O}}}(\bar{E},\omega)$.  To proceed any further, we must again specify what sorts of local operators $\hat{\mathcal{O}}^{(A)}$ we wish to consider. We again consider `typical' operators on the $A$ factor (see equation \eqref{eq: RMT result}), replacing $|\hat{\mathcal{O}}^{(A)}_{ij}|^2 \rightarrow |\mathcal{H}^{(A)}|^{-1}\overline{\mathcal{O}^{2(A)}}$, where $\overline{\mathcal{O}^{2(A)}}=\overline{\mathcal{O}^{2}}$ represents the average of the operator's squared spectrum.  If we now also restrict to a regime where $n_B$ is approximately constant between $\bar{E}-\overline{E^{(A)}}\pm\sigma_A/2$ (which in our examples will just mean taking small subsystems, $\sigma_A<<\sigma_B$), then this simplifies to
\begin{equation}\begin{split}\label{eq: f off-diag, smooth scrambling}
    f_{\hat{\mathcal{O}}}(\bar{E},\omega)
    &\approx
    \sqrt{\frac{\overline{O^2}\sigma_S}{N_h^2 |\mathcal{H}^{(A)}|^2}\sum_{ij}[h\star h](2\omega-2\omega^{(A)}_{ij})}
    \approx
    \sqrt{2\overline{\mathcal{O}^2}
    \frac{\sigma_S}{N_h^2} \int \dd\omega'~ 
    [\rho_A\star\rho_A](2\omega')[h\star h](2\omega-2\omega')}.\\
\end{split}\end{equation}
The last expression switches to a continuum approximation of the remaining sums.
Recall that $[g_1\star g_2](x):=\int \dd y~ g_1(y)g_2(y+x)$ denotes a cross-correlation and $\rho_A:=|\mathcal{H}^{(A)}|^{-1}n_A$ denotes the fractional density of states of $\hat{H}^{(A)}$ (such that $\int\dd\epsilon~ \rho_A(\epsilon) = 1$). Equation \eqref{eq: f off-diag, smooth scrambling} corrects the result of equation \eqref{eq: f narrow scrambling small A limit} to incorporate affects of finite width and nontrivial tails in the scrambling ansatz $h$ (equation \eqref{eq: microcanonical, narrow scrambling limit,  appendix} in the appendix is just \eqref{eq: f off-diag, smooth scrambling} applied to the `perfect microcanonical scrambling' ansatz).
To get a feeling for the contributions under the integral in the last expression, we may consider a variety of functional forms of the scrambling ansatz $h(E)$, and likewise the density of states $\rho_A$.

A scrambling ansatz that seems to be well motivated numerically is to take the variance to decay exponentially around the target energy, by which we mean taking
\begin{equation}\begin{split}\label{eq: h exp decay}
    h(E)&:=\exp(\frac{-\sqrt{2}|E|}{\sigma_S}),\\
    N_h &= \sqrt{2}\sigma_S,\\
    [h\star h](E)&=(\sigma_S/\sqrt{2}+|E|)~ \exp(\frac{-\sqrt{2}|E|}{\sigma_S}).
\end{split}\end{equation}
For small subsystems $A$, a reasonable first approximation to the density of states $\rho_A$ is to take a flat distribution over the spectral range $\sigma_A$.  In this case, $[\rho\star \rho](2\omega')$ becomes $\frac{1}{\sigma_A}\left(1-\frac{2|\omega'|}{\sigma_A}\right)\Theta\left(1-\frac{2|\omega'|}{\sigma_A}\right)$. Employing this choice along with \eqref{eq: h exp decay} in \eqref{eq: f off-diag, smooth scrambling} leads to
\begin{equation}\begin{split}\label{eq: off-diag exp decay}
    f_{\hat{\mathcal{O}}}(\bar{E},\omega)\approx
    \sqrt{\frac{\overline{\mathcal{O}^{2}}}{2\sqrt{2}}
   \int_{-1}^{1}\dd x (1-|x|)\left(1+\frac{\sqrt{2}|2\omega-x\sigma_A|}{\sigma_S}\right)\exp(\frac{-\sqrt{2}|2\omega-x\sigma_A|}{\sigma_S})}.
\end{split}\end{equation}
Here we have switched to a dimensionless variable $\omega\rightarrow x\sigma_A/2$ for the remaining integration.  We will compare equation \eqref{eq: off-diag exp decay} to results in a chaotic spin chain system in the following section.

\section{Numerical illustrations}
\label{sec: numerical illustrations}
\subsection{1-D chaotic spin chain}
\label{subsec: 1-D spin chain}

\begin{figure}[H]
		\centering
		\includegraphics[width=.5\linewidth]{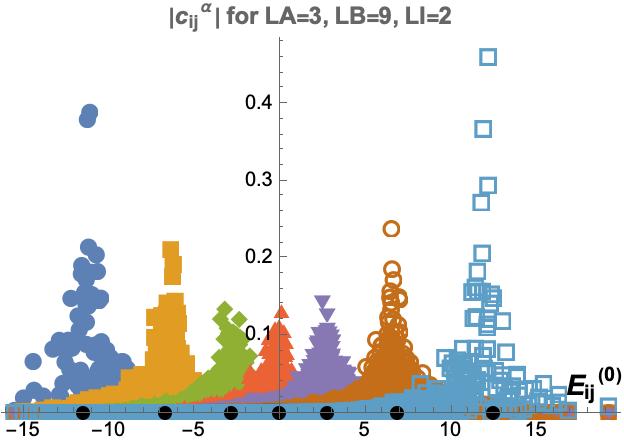}
	\caption{Coefficient magnitudes are shown for eigenstates of $\hat{H}^{(T)}$ of equation \eqref{eq: qubit hamiltonian} when expressed in the eigenbasis of $\hat{H}^{(0)}$, which is the same Hamiltonian with a single interaction term removed.  The case shown is for $L_A=3$ and $L_B=9$, so the relevant interaction term is $H^{(I)}=J \sigma^{3}_z\sigma^{4}_z$.  Seven different eigenvalues ${E^{(T)}_\alpha}$ of the total system are chosen at energies spread across the spectrum, represented by the black dots on the horizontal axis.  The magnitudes of coefficients $|c^{\alpha}_{ij}|:=|\braket*{E^{(T)}_\alpha}{E^{(A)}_{i},E^{(B)}_{j}}|$ for all $i, j$ are plotted in different colors for each of the seven energies.  The qualitative clustering on energy windows is evident, with the variance of scrambling found to be $\sigma_S\sim 0.96$ toward the center of the spectrum.}
	\label{fig: coefficient scrambling ising}
\end{figure}

We now consider a one-dimensional spin chain system that is well-studied in the context of quantum chaos \cite{Banuls:2011vuw,Shenker:2013pqa,Hosur:2015ylk,Couch:2019zni,Eccles:2021zum}:
\begin{equation}\begin{split}\label{eq: qubit hamiltonian}
		\hat{H}^{(T)} &= J\sum_{r=1}^{L-1}\sigma_{z}^{(r)} \sigma_{z}^{(r+1)}
		+h_x\sum_{r=1}^{L}\sigma_{x}^{(r)}  +h_z\sum_{r=1}^{L}\sigma_{z}^{(r)}. \\
\end{split}\end{equation}
Here, $\sigma_i^{(r)}$ denotes a Pauli operator on the $r$'th site, for $i=\{x,y,z\}$.  For coupling strengths $J=1, h_x=1.05,$ and $h_z=0.5$, which we employ throughout, the Hamiltonian \eqref{eq: qubit hamiltonian} is known to be highly chaotic, as diagnosed through level-spacing statistics and related criteria.  We will take advantage of the pre-given tensor product structure and consider operators localized to the first $L_A$ factors, for $L_A = \{1,2,3,...\}$. The interaction term in each case is then simply $J \sigma^{L_A}_z\sigma^{L_A+1}_z$, with a corresponding Hamiltonian decomposition and perturbation problem (see equation \eqref{eq: hamiltonian decomposition}).  
We first test the expectation that the associated perturbation problem results in localized eigenstate scrambling.  Figure \ref{fig: coefficient scrambling ising} illustrates that this is indeed the case, using $L_A=3$ in a 12 qubit system as an example.  The distribution of support is shown for a handful of different eigenstates of $\hat{H}^{(T)}$ in the $\hat{H}^{(0)}$ eigenbasis.  The scrambling variance $\sigma_S$ of these distributions is found to be very close to $1$ for $E^{(T)}_\alpha$ far from the edges of the spectrum.  This matches our expectation that the scrambling variance is on the order of $||\hat{H}^{(I)}||$, which in this case is precisely $1$, though we leave it as a matter for future work to understand the general aspects of the perturbation problem that control the scrambling statistics.

Next, to test the behavior of operators localized on subfactors of this system, we generate a collection of such operators and average their norm squared values as functions of $\bar{E}=(E^{(T)}_\alpha+E^{(T)}_\beta)/2$ and $\omega=(E^{(T)}_\alpha-E^{(T)}_\beta)/2$ for comparison with the predictions of the scrambling ansatzes of sections \ref{subsec: microcanonical ansatz} and \ref{subsec: smooth ansatz}. Our procedure is as follows. A random spectrum of $2^{L_A}$ eigenvalues from a flat distribution between -1 and 1 is selected and then normalized to have mean zero and have average squared spectrum $\overline{O^2}=1$.  This is then localized as a diagonal operator on the $A$ factor, after which point a random orthogonal transformation on the $A$ factor is applied. From the resulting operator, the magnitude squared of matrix elements are obtained in the total energy eigenbasis.  These are binned into $\omega$ bins of size $\Delta\omega\sim 0.015$, and energy bins of $\bar{E}\pm 1/2$ around a target energy $\bar{E}$.  The same process is repeated for $250$ random operators on the same factor, and then the average magnitude squared of off-diagonal matrix elements as a function of $\bar{E}$ and $\omega$ are compared against the predictions of the scrambling ansatzes of sections sections \ref{subsec: microcanonical ansatz} and \ref{subsec: smooth ansatz} in various forms.  Results are shown in figures \ref{fig: off-diag, scan LA}  and \ref{fig: off-diag, scan E}, which we now discuss in detail.

Figure \ref{fig: off-diag, scan LA} shows results for operators across subsystem sizes $L_A=1,3,5,7$ in a 12-qubit system.  Dots represent average norm squared matrix elements of random local operators, collected and binned as described above.  The solid lines represent the result of equation \eqref{eq: smooth result off diagonal} for typical operators $|\hat{\mathcal{O}}_{ij}|^2\rightarrow \overline{\mathcal{O}^2}/|\mathcal{H}_A|$, using the scrambling ansatz \eqref{eq: h exp decay} (motivated by distributions of the sort shown in figure \ref{fig: coefficient scrambling ising}).  It can be seen that the curves provide a very good characterization of the scale and the falloff as a function of $|\omega|$, though the fit is less accurate as $|\omega|\rightarrow 0$. In particular, the curves seem to miss a peak feature that generically occurs in the data as $|\omega|\rightarrow 0$.  We take this as a possible indication that the assumption of completely independent (uncorrelated) coefficients $c^{\alpha}_{ij}$, $c^{\beta}_{ij}$ for very nearby $\alpha, \beta$, may be unwarranted.  Regardless, this peak feature becomes less prominent as $L_A$ increases.  In the left panel, all curves are at $\bar{E}=0$, which is the center of the spectrum.  Here the approximation \eqref{eq: smooth result off diagonal} should be most robust due to assumptions about slowly varying density of states $n_B$.  The right panel shows results halfway between the edge and center of the spectrum at $\bar{E}=0.5 E^{(T)}_{\text{min}} \approx -7.85$.  These show larger variance in the data but still a good fit.  Note the change of vertical scales between left and right figures, which is largely associated with the differing entropic suppression factors $e^{-S(\bar{E})/2}$.

Figure \ref{fig: off-diag, scan E} shows results for a fixed subsystem size $L_A$ as $\bar{E}$ is varied between $0$ and $0.5\bar{E}_{\text{min}}\approx -7.85$.  In the left diagram, the solid lines represent approximation \eqref{eq: off-diag exp decay}, which eliminates dependence on spectral density interpolations (except in the entropic suppression factor) by crudely approximating the small, discrete spectrum of $\hat{H}^{(A)}$ with a flat spectral density, and assuming $\sigma_B>>\sigma_A,\sigma_S$.  In the right diagram, the solid lines represent approximation \eqref{eq: narrow scrambling limit}, which is a ``narrow scrambling'' limit that eliminates all dependence on the finite width or detailed form of the scrambling ansatz.  In both diagrams, it would be more accurate to employ equation \eqref{eq: smooth result off diagonal} or even \eqref{eq: discrete result} with exact sums, which is still tractable in a $12$ qubit system.  Here we employ these approximations merely to illustrate that such simplifications still capture the essential behavior.  For systems larger than the 12 qubits we consider, such approximations become both more necessary and more accurate.  In particular, in systems with local interactions such as this, where the scrambling scale grows more slowly (or not at all) as the scales $\sigma_A$ and $\sigma_B$ increase in fixed ratio, the narrow scrambling approximation is particularly relevant.

\begin{figure}[!htb]
\minipage{0.52\textwidth}
\centering
  \includegraphics[width=\linewidth]{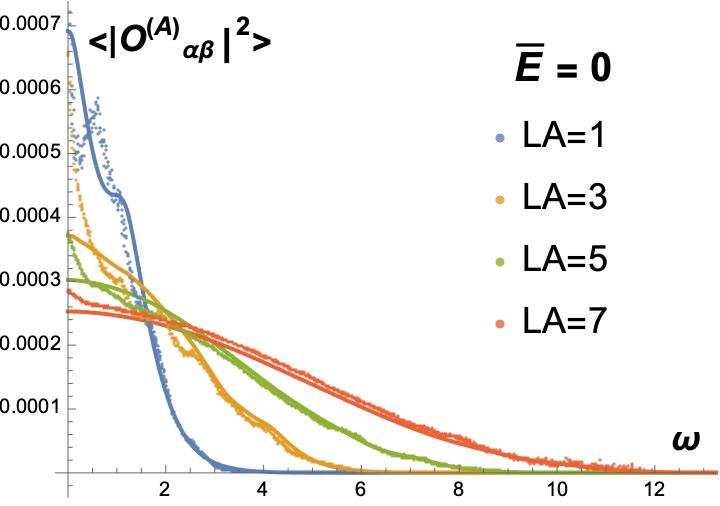}
\endminipage\hfill
\minipage{0.48\textwidth}
\centering
  \includegraphics[width=\linewidth]{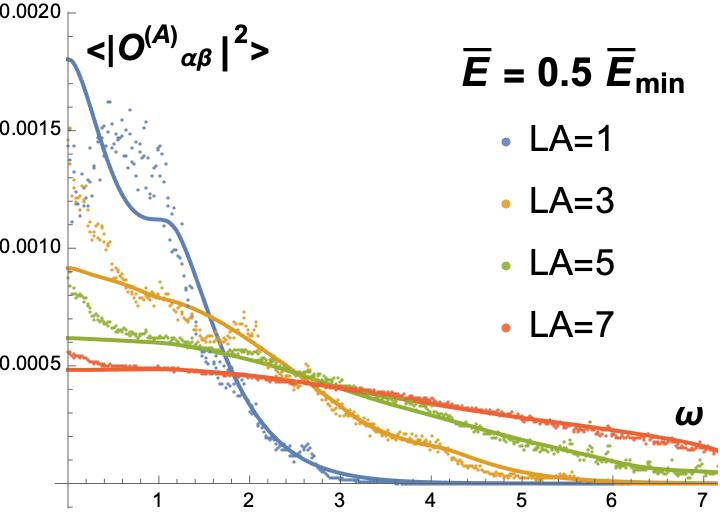}
\endminipage\hfill
\caption{Average squared magnitudes of matrix elements are shown for typical operators localized on the first $L_A$ qubit factors of the system \eqref{eq: qubit hamiltonian}.  Solid lines represent the prediction of equation \eqref{eq: smooth result off diagonal} for typical operators ($|\hat{\mathcal{O}}_{ij}|^2\rightarrow \overline{\mathcal{O}^2}/|\mathcal{H}_A|$) using interpolated density of states functions and scrambling ansatz \eqref{eq: h exp decay}.  Mirror image results occur for $\omega<0$.  See main text for discussion.}
\label{fig: off-diag, scan LA}
\end{figure}

\begin{figure}[!htb]
\minipage{0.5\textwidth}
\centering
  \includegraphics[width=\linewidth]{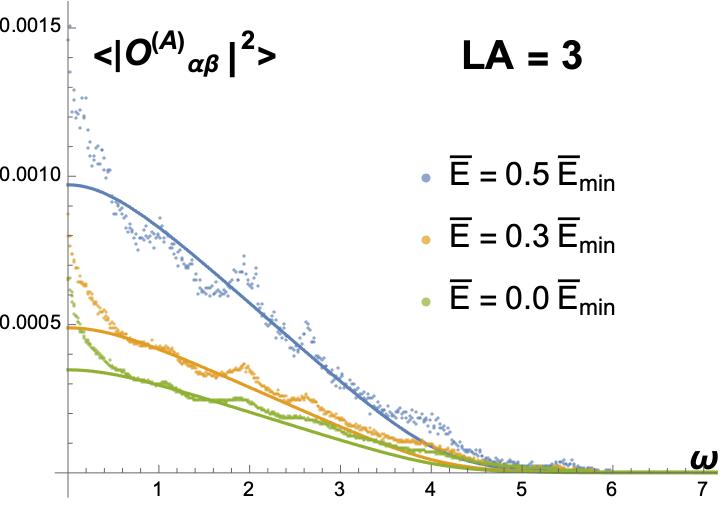}
\endminipage\hfill
\minipage{0.5\textwidth}
\centering
  \includegraphics[width=\linewidth]{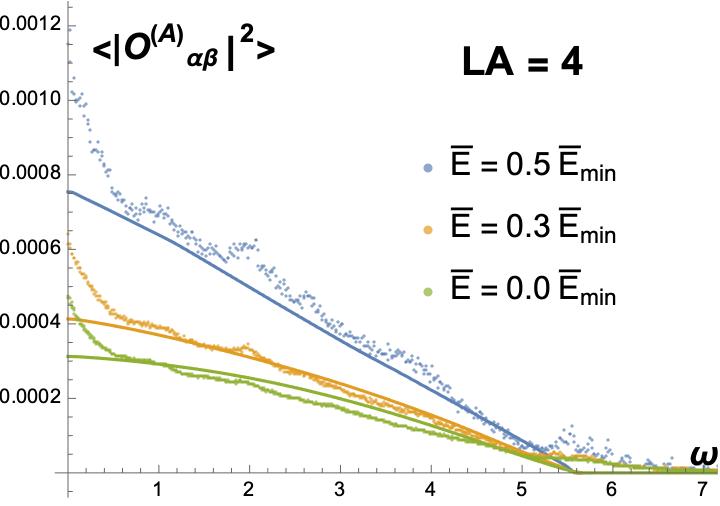}
\endminipage\hfill
\caption{Each plot shows, for fixed $L_A$, the $\omega$ dependence of matrix elements at three different energies $\bar{E}$, ranging from the center of the spectrum ($\bar{E}=0$) to halfway between the center and edge of the spectrum ($\bar{E}=0.5 \bar{E}_{\text{min}} \approx -7.85$).  Although the exact sums of equation \eqref{eq: smooth result off diagonal} would be more accurate, the solid lines in these plots represent the approximations of equation \eqref{eq: off-diag exp decay} (left), and equation \eqref{eq: narrow scrambling limit} (right). Mirror image results occur for $\omega<0$. See main text for discussion.}
\label{fig: off-diag, scan E}
\end{figure}

In these tests and illustrations, we have focused on ``typical'' local operators.  For other individual operators, the general expressions of equation \eqref{eq: smooth result off diagonal} will of course be more accurate; the detailed form of the $f_{\hat{\mathcal{O}}}$ will include different features, but the overall scale and width of the falloff are captured already by considerign typical operators.  In this sense, the expressions for typical operators provide a characterization of the entire algebra of local operators on the $\hat{H}^{(A)}$ factor.
 
\section{Discussion and future directions}
\label{sec: discussion and future directions}

In this work we have studied the interplay of operator locality with the ETH ansatz for operator matrix elements.  We introduced an operator-specific perturbation problem that elucidates the circumstances and extent to which specific operators are of `ETH-class' in a given system.  Of course, the implication is not that every single operator should be separately subjected to similar analysis to determine whether or not ETH applies.  Rather, the framework provides a means of conceptualizing which aspects of the operator-system relationship are important to the ETH ansatz, and to thermalization in general, in a manner that allows for distinct behavior among different classes of operators in the same system.  We avoid blanket statements about the onset of ETH for \textit{all} operators in a given system (which would be manifestly incorrect), and even more appropriate statements about all \textit{local} operators in a given system without being explicit about the relevant notion of locality (since different tensor product structures might be fruitfully invoked even in the same system). We study the implications of a few simple scrambling ansatzes for the relevant perturbation problem, deriving the off-diagonal behavior of typical local operators in various regimes.  The spectral properties of the part of the Hamiltonian that acts on the same factor as the chosen operator is shown to play an important role in the details of the off-diagonal ETH ansatz, setting the scale over which the (already suppressed) off-diagonal matrix elements decay to zero in locally interacting systems.  In fact, the behavior of typical operators among the same factor algebra can be thought of as providing a partial charaterization of this subsystem Hamiltonian (see equation \eqref{eq: f narrow scrambling small A limit}, for instance, and also appendix \ref{app: random hamiltonian systems}).

As alluded to already, although we have referred to the above perturbation problem as `operator-specific', the same decomposition is relevant to the thermalization of all operators local to the same tensor factor.  For this reason it might better be referred to as an `algebra-specific' perturbation problem.  Strictly speaking, a perturbation problem which is even more relevant to the thermalization of an \textit{individual} operator is one which projects the Hamiltonian onto all of the operator eigenspaces, separating a piece which commutes with the operator and a remainer which doesn't.  Such projections are rightfully at the heart of a lot of ETH-related analysis, but since the resulting decomposition does not generically coincide with any factorization, it is not equally amenable certain standard techniques of open-subsystem analysis.  We consider the focus on local operators to present an optimal combination of physical relevence, intuitive clarity, and analytical tractability.

In the future it would be interesting to study both the `generic' perturbation problem (see appendix \ref{app: random hamiltonian systems}) as well as the specific perturbation problem as it arises in systems that exhibit unusual thermalization properties (e.g. many body-localizing systems).  In principle, all standard techniques of open system analysis may be brought to bear on the generic problem.  Although we have focused exclusively on operator matrix elements, any given scrambling ansatz also has implications for the entanglement entropy on the subfactor $A$, which would be interesting to explore.  It would also be natural to generalize scrambling ansatzes to cases when conserved charges are present, which inevitably entails more structured eigenstate scrambling.  Such structured ansatzes would inevitably make contact with recent work such as \cite{PhysRevLett.130.140402} and \cite{Bianchi_2024}.  All these directions we leave for future work.

\acknowledgments

We thank Elena Caceres, Nick Hunter-Jones, Jason Pollack, Sarah Racz, Suvrat Raju, Nilakash Sorokhaibam, and the whole of the Qubits and Spacetime Unit at Okinawa Institute of Science and Technology for helpful comments and discussion.  
We thank the International Centre for Theoretical Sciences,  Bangalore, as well as the Aspen Center for Physics for hosting visits over which part of this work was completed.
The work was supported by funding from the Okinawa Institute of Science and Technology Graduate University.~It was also made possible through the support of the ID\# 62312 grant from the John Templeton Foundation, as part of the \href{https://www.templeton.org/grant/the-quantum-information-structure-of-spacetime-qiss-second-phase}{\textit{`The Quantum Information Structure of Spacetime'} Project (QISS)} from the John Templeton Foundation.~The opinions expressed in this project/publication are those of the author(s) and do not necessarily reflect the views of the John Templeton Foundation..

\appendix

\section{Approximations from perfect microcanonical scrambling}
\label{app: approximations from perfect microcanonical scrambling}
Here we work out a few implications of the ``perfect microcanonical scrambling'' ansatz of section \ref{subsec: microcanonical ansatz - off diagonal}.
We always work in a regime where the scrambling width $\Delta$ is narrow compared to the scale on which $n_B$ changes, which allows us to approximate the relevant sub-Hilbert space dimensions via
\begin{equation}\begin{split}\label{eq: sub hilbert space approx}
    |\mathcal{H}^{(B)}_{E_1}| &= \int_{E_1-\Delta/2}^{E_1+\Delta/2}\dd\epsilon~ n_B(\epsilon)\approx
    n_B(E_1)\Delta,\\
    |\mathcal{H}^{(B)}_{E_1}\cap\mathcal{H}^{(B)}_{E_2}|
    &=\int_{\max(E_1,E_2)-\Delta/2}^{\min(E_1,E_2)+\Delta/2}\dd\epsilon~n_B(\epsilon)
    \approx
    n_B\left(\bar{E}_{12}\right)(\Delta-2|\omega_{12}|)\Theta(\Delta-2|\omega_{12}|),
\end{split}\end{equation}
where $\Theta$ is a Heaviside step function. These and related approximations below also require that we avoid arguments (particularly $\bar{E}$) near the edges of the full spectrum.  We then have
\begin{equation}\begin{split}\label{eq: sum approximations 0, appendix}
    \sum_{i,j}&\big|\mathcal{H}^{(B)}_{E^{(T)}_\alpha-E^{(A)}_i}\cap\mathcal{H}^{(B)}_{E^{(T)}_\beta-E^{(A)}_j}\big|
    |\hat{\mathcal{O}}^{(A)}_{ij}\big|^2
    \approx
    \sum_{i,j}n_B\left(\bar{E}^{(T)}_{\alpha\beta}-\bar{E}^{(A)}_{ij}\right)(\Delta-2|\omega^{(T)}_{\alpha\beta}-\omega^{(A)}_{ij}|)\Theta[\Delta-2|\omega^{(T)}_{\alpha\beta}-\omega^{(A)}_{ij}|]\big|
    |\hat{\mathcal{O}}^{(A)}_{ij}\big|^2\\
    &\approx
    2\int \dd\epsilon \int\dd\omega ~n_A(\epsilon+\omega)n_A(\epsilon-\omega)n_B(\bar{E}^{(T)}_{\alpha\beta}-\epsilon)(\Delta-2|\omega^{(T)}_{\alpha\beta}-\omega|)\Theta[\Delta-2|\omega^{(T)}_{\alpha\beta}-\omega|]\big|\bra{\epsilon+\omega}\hat{\mathcal{O}}^{(A)}\ket{\epsilon-\omega}\big|^2.
\end{split}\end{equation}
In the second line we've switched to continuum approximations for the remaining sums (which requires that the operator $\hat{\mathcal{O}}^{(A)}$ be smoothed/smeared accordingly).  Integrations can be taken over the full real line with the understanding that the spectral densities vanish outside a bounded spectral range.  A variety of approximations may now be considered.  In a `narrow scrambling' limit, where the scrambling width $\Delta$ is small compared to the spectral range $\sigma_A$ (regardless of the comparative size of $A$ and $B$ factors), the leading behavior of \eqref{eq: sum approximations 0, appendix} is
\begin{equation}\begin{split}\label{eq: sum approximations 1, appendix}
    \Delta^2 \int \dd\epsilon ~n_A(\epsilon+\omega^{(T)}_{\alpha\beta})n_A(\epsilon-\omega^{(T)}_{\alpha\beta})
    n_B(\bar{E}^{(T)}_{\alpha\beta}-\epsilon)
    ~\big|\bra*{\epsilon+\omega^{(T)}_{\alpha\beta}}\hat{\mathcal{O}}^{(A)}\ket*{\epsilon-\omega^{(T)}_{\alpha\beta}}\big|^2, \quad \sigma_S<<\sigma_A,\sigma_B
\end{split}\end{equation}
If we replace $\Delta$ with the corresponding variance $\sigma_S$ of the flat scrambling distribution $(\Delta=2\sqrt(3)\sigma_S)$, and define a microcanonical suppression factor of $e^{-S(\bar{E}^{(T)}_{\alpha\beta})/2}:=(\sigma_S n_0(\bar{E}^{(T)}_{\alpha\beta}))^{-1/2}$, this leads to
\begin{equation}\begin{split}\label{eq: narrow scrambling limit, full op, appendix}
    f_{\hat{\mathcal{O}}}(\bar{E},\omega)&\approx
    \sqrt{\frac{\sigma_S n_0(\bar{E})}{n_0(\bar{E}+\omega)n_0(\bar{E}-\omega)}\int \dd\epsilon ~n_A(\epsilon+\omega)n_A(\epsilon-\omega)
    n_B(\bar{E}-\epsilon)|\bra*{\epsilon+\omega}\hat{\mathcal{O}}^{(A)}\ket*{\epsilon-\omega}|^2},\quad \sigma_S<<\sigma_A,\sigma_B.
\end{split}\end{equation}
If we further restrict to the case that the density $n_B$ is effectively constant in the range $[\bar{E}-\overline{E^{(A)}}-\sigma_A/2,\bar{E}-\overline{E^{(A)}}+\sigma_A/2]$ (which for simple systems means the regime $\sigma_S<<\sigma_A<<\sigma_B$) this becomes
\begin{equation}\begin{split}\label{eq: narrow scrambling small A limit, full op, appendix}
    f_{\hat{\mathcal{O}}}(\bar{E},\omega)&\approx
    \sqrt{\frac{\sigma_S }{|\hat{H}^{(A)}|}\int \dd\epsilon ~n_A(\epsilon+\omega)n_A(\epsilon-\omega)|\bra*{\epsilon+\omega}\hat{\mathcal{O}}^{(A)}\ket*{\epsilon-\omega}|^2},\quad \sigma_S<<\sigma_A<<\sigma_B.
\end{split}\end{equation}
For `typical operators' on the $A$ factor, which corresponds to making the replacement $|\hat{\mathcal{O}}^{(A)}_{ij}|^2 \rightarrow |\mathcal{H}^{(A)}|^{-1}\overline{\mathcal{O}^{2(A)}}$, with $\overline{\mathcal{O}^{2(A)}}=\overline{\mathcal{O}^{2}}$ being the average of the squared operator spectrum, equations \eqref{eq: narrow scrambling limit, full op, appendix} and \eqref{eq: narrow scrambling small A limit, full op, appendix} become, respectively,
\begin{equation}\begin{split}\label{eq: narrow scrambling limit, typical op, appendix}
    f_{\hat{\mathcal{O}}}(\bar{E},\omega)&\approx
    \sqrt{ \overline{\mathcal{O}^2}\frac{\sigma_S\rho_0(\bar{E})}{\rho_0(\bar{E}+\omega)\rho_0(\bar{E}-\omega)}\int \dd\epsilon ~\rho_A(\epsilon+\omega)\rho_A(\epsilon-\omega)
    \rho_B(\bar{E}-\epsilon)},\quad \sigma_S<<\sigma_A,\sigma_B,
\end{split}\end{equation}
and
\begin{equation}\begin{split}\label{eq: narrow scrambling small A limit, typical op, appendix}
    f_{\hat{\mathcal{O}}}(\bar{E},\omega)&\approx
    \sqrt{\sigma_S\overline{\mathcal{O}^2} \int \dd\epsilon ~\rho_A(\epsilon+\omega)\rho_A(\epsilon-\omega)},\quad \sigma_S<<\sigma_A<<\sigma_B.
\end{split}\end{equation}
In both expressions we've switched to the fractional density of states, e.g. $\rho_A = n_A |\mathcal{H}^{(A)}|$.  Note that \eqref{eq: narrow scrambling small A limit, typical op, appendix} has become independent of $\bar{E}$, though matrix elements still depend on $\bar{E}$ through the entropic suppression term.
We may also consider a regime $\sigma_S,\sigma_A<<\sigma_B$ which more accurately captures the finite width of the scrambling window by returning to equation \eqref{eq: sum approximations 0, appendix} and considering only that $n_B$ is essentially constant in the integral over the spectral range of $\hat{H}^{(A)}$.  For typical operators, this leads to 
\begin{equation}\begin{split}\label{eq: microcanonical, narrow scrambling limit,  appendix}
    f_{\hat{\mathcal{O}}}(\bar{E},\omega)&\approx
    \sqrt{\frac{\overline{\mathcal{O}^2}\sigma_A}{2\sqrt{3}}\int_{-1}^{1}\dd x ~[\rho_A\star \rho_A](x\sigma_A)\left(1-\frac{|\omega-x\sigma_A/2|}{\sqrt{3}\sigma_S}\right)\Theta\left(1-\frac{|\omega-x\sigma_A/2|}{\sqrt{3}\sigma_S}\right)}, \quad \sigma_S,\sigma_A<<\sigma_B,
\end{split}\end{equation}
where we denote the auto-correlation $[\rho_A\star\rho_A](2\omega):=\int\dd\epsilon\rho_A(\epsilon)\rho_A(\epsilon+2\omega)$, and we've changed to a dimensionless integration variable $\omega\rightarrow x\sigma_A/2$. This is a `perfect microcanonical scrambling' analogue of \eqref{eq: f off-diag, smooth scrambling} for more realistic scrambling ansatzes.  Note that in the narrow scrambling limit $\sigma_S\rightarrow 0$ it reduces again to \eqref{eq: narrow scrambling small A limit, typical op, appendix} for $\sigma_S<<\sigma_A<<\sigma_B$.

\section{Random Hamiltonian systems}
\label{app: random hamiltonian systems}
We now describe a setup for constructing random Hamiltonians that can be used to study the `generic' perturbation problem of section \ref{subsec: operator-specific perturbation}.  Here we only use it to numerically illustrate some further features of off-diagonal matrix elements that can occur in a non-ETH regime, which were briefly described in section \ref{subsec: microcanonical ansatz - off diagonal}.

\begin{figure}[H]
		\centering
		\includegraphics[width=.5\linewidth]{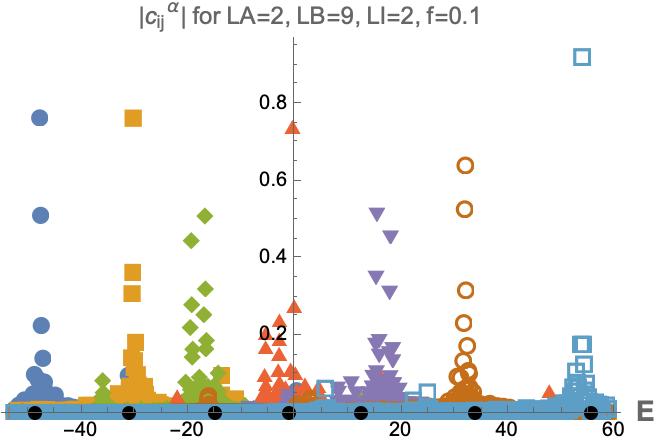}
	\caption{Coefficient magnitudes are shown for eigenstates of the $\hat{H}^{(T)}$ system expressed in the $\hat{H}^{(0)}$ eigenbasis for the random Hamiltonian system as described in the main text.}
	\label{fig: random Hamiltonian coeff scrambling}
\end{figure}

We start by generating two random (GOE) Hamiltonians to serve as $\hat{H}^{(A)}$ and $\hat{H}^{(B)}$.  These are given the dimension of an $L_{A}$ qubit Hilbert space and $L_{B}$ qubit Hilbert space, respectively.  The unperturbed Hamiltonian is then constructed as $\hat{H}^{(0)} = \hat{H}^{(A)}\otimes \mathbb{I}^{(B)} +  \mathbb{I}^{(A)}\otimes \hat{H}^{(B)}$, with the dimension of an $(L_A+L_B)$-qubit Hilbert space.  An interaction term is then generated, which we here take to be another GOE Hamiltonian, acting on an $L_I$-qubit subfactor which spans the $A$ and $B$ factors.\footnote{More precisely, we generate only the GOE spectra of $\hat{H}^{(A)}$ and $\hat{H}^{(B)}$, placing these in a diagonal matrix for $\hat{H}^{(0)}$. A coupling term of the form $O_{L_A}\otimes O_{L_B} \left(\mathbb{I}_{L_{A}-\text{floor}(L_I/2)} \otimes \hat{H}_{L_I} \otimes \mathbb{I}_{L_B-\text{ceiling}(L_I/2)}\right) O^\dagger_{L_A}\otimes O^\dagger_{L_B}$ is then added, where $\hat{H}_{L_I}$ is an $L_I$-qubit diagonal Hamiltonian with GOE spectrum, and $O_{L_A}$ and $O_{L_B}$ are random orthogonal matrices on $\mathcal{H}^{(A)}$ and $\mathcal{H}^{(B)}$, respectively.}  The operator norm of the interaction term is scaled to a chosen parameter $f:=||\hat{H}^{(I)}||/||\hat{H}^{(0)}||$, quantifying the strength of the interaction.  Figure \ref{fig: random Hamiltonian coeff scrambling} show the localized scrambling of eigenstates from one such system, though the coefficient distributions can be differ drastically depending on various choices in the instantiation of random spectra for the Hamiltonians $\hat{H}^{(A)}$, $\hat{H}^{(B)}$, and $\hat{H}^{(I)}$.

\begin{figure}[!htb]
\minipage{0.5\textwidth}
\centering
  \includegraphics[width=\linewidth]{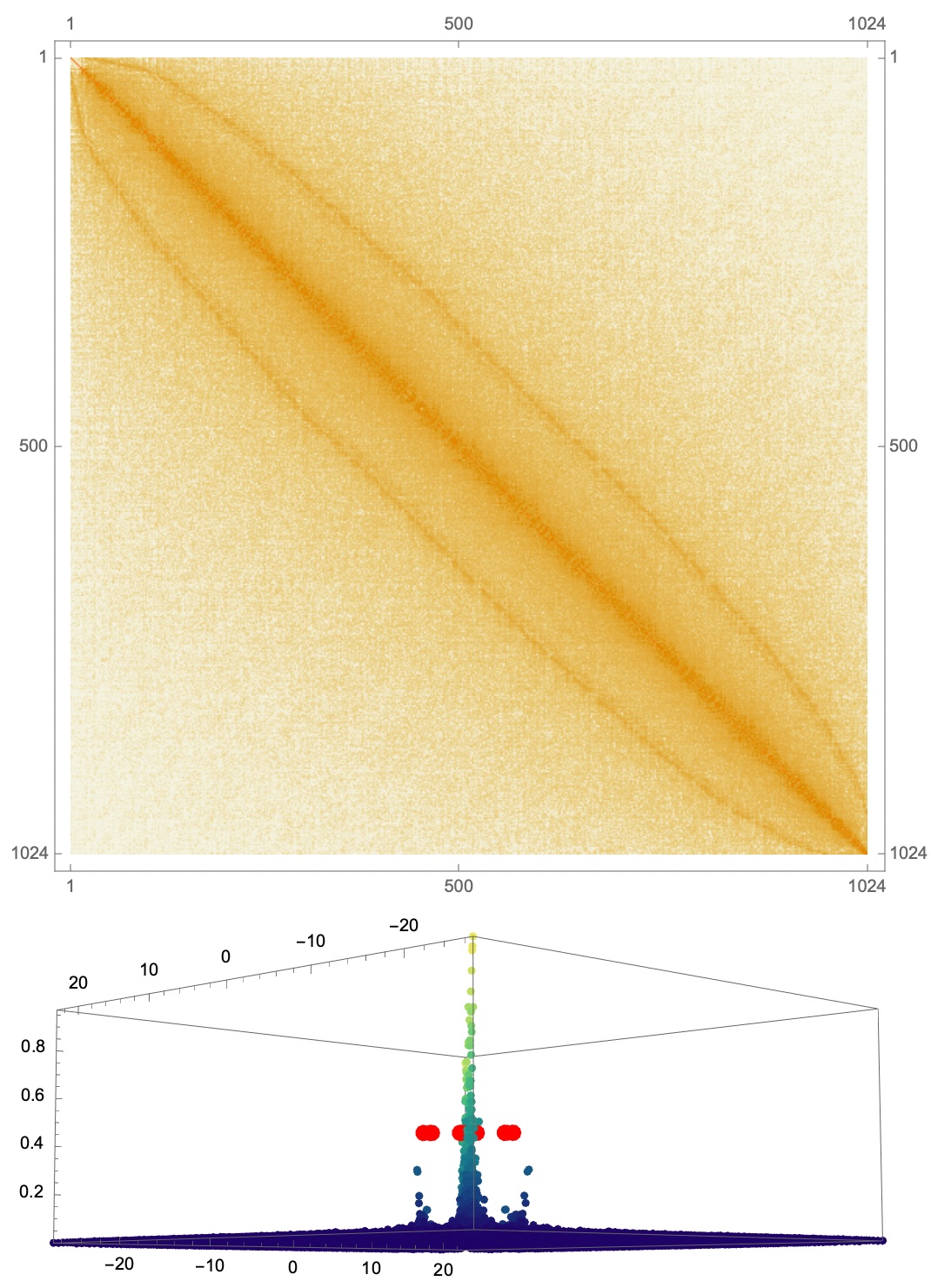}
\endminipage\hfill
\minipage{0.5\textwidth}
\centering
  \includegraphics[width=\linewidth]{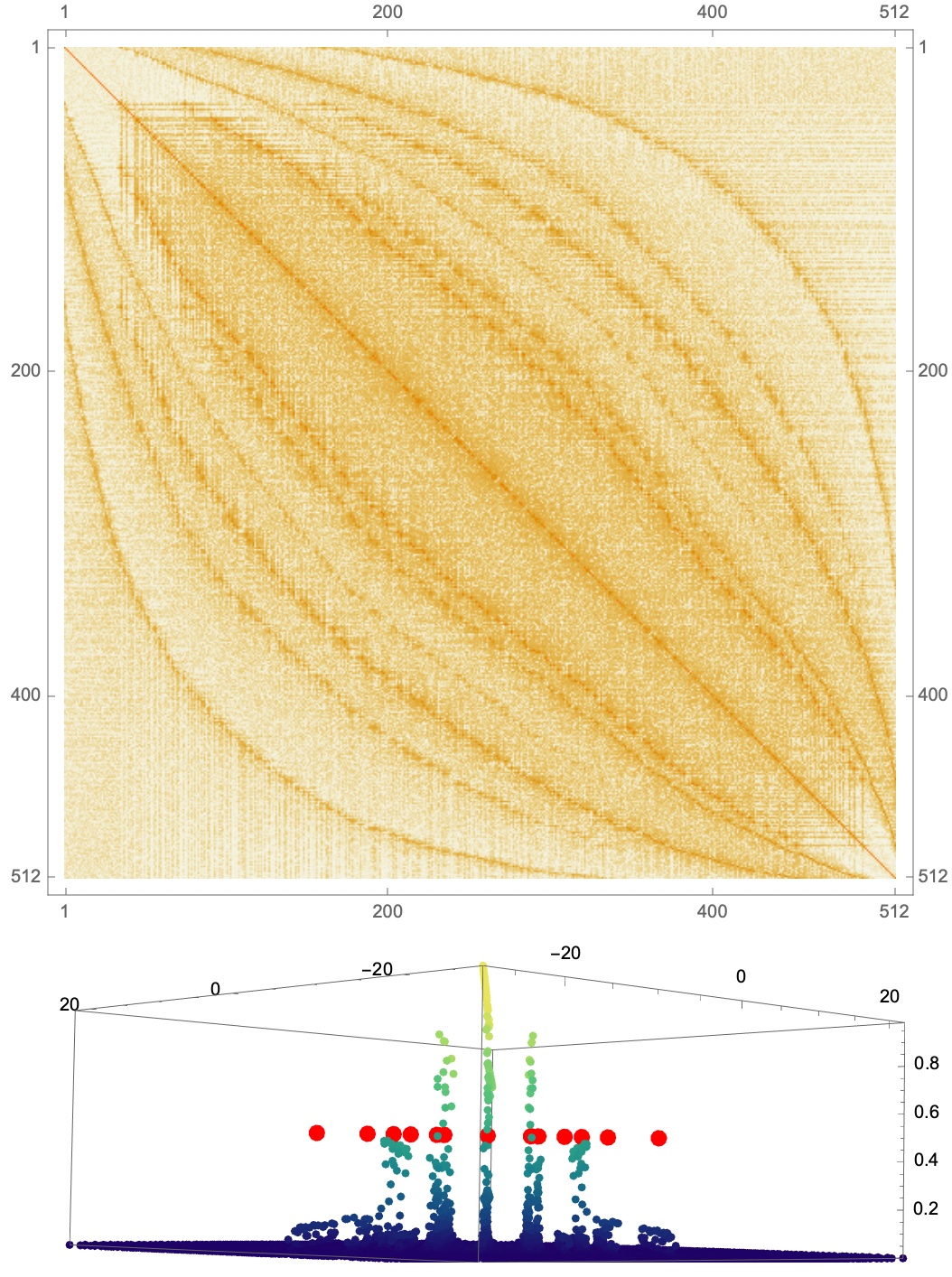}
\endminipage\hfill
\caption{The magnitudes of matrix elements are displayed for operators localized on an $L_A=2$ qubit factor of a random Hamiltonian system as described in the main text. The upper plots are full matrix plot, while the lower plots display the same matrix in a 3D scatter plot looking ``down the diagonal'' to see banding as a function of $\omega=(E^{(T)}_\alpha-E^{(T)}_\beta)/2$.  Red dots are placed in positons corresponding to the spectral gaps of the $\hat{H}^{(A)}$ Hamiltonian.  See main text for discussion.}
\label{fig: banded matrices}
\end{figure}

This or a similar setup may be used to study the `generic' perturbation problem of equation \ref{subsec: operator-specific perturbation} to understand how the statistics of microstate scrambling statistics depend on features of the decomposed Hamiltonian(s) and their interplay.  Here, however, we only use it to illustrate a specific regime that was described in section \ref{subsec: microcanonical ansatz - off diagonal}, where the scale of the scrambling is chosen to be on the same or smaller scale than the spectral gaps of the $\hat{H}^{(A)}$ Hamiltonian. In such cases, these spectral gaps can imprint a banding structure on the off-diagonal matrix elements of operators localized on the same factor.  This is illustrated in figure \ref{fig: banded matrices} for a system of $L_A=2$ and $L_B=9$.  Both figures show matrix plots of the magnitudes of $L_A$-local operators in the total energy eigenbasis, as well as 3D scatter plots looking `down the diagonal' on the same matrix elements.  Banding is apparent already in the matrix plots, but since the axes are indexical values of the sorted eigenbasis rather than the energy itself, the bands do not appear linear.  In the scatter plots the horizontal axes correspond to energy, so this diagonal perspective makes it apparent that the banding is essentially a function of $\omega$ alone.  The red dots are placed at $\omega$ values corresponding to the spectral gaps of the $\hat{H}^{(A)}$ Hamiltonian. Their vertical placement is arbitrary (chosen for clarity).  It can be seen that these spectral gaps are in approximate correspondance to the banding structure in the of matrix elements (although the full details of the bands also depend on the randomly selected operator and the interaction Hamiltonian).  In the right plots, the $L_A$ Hamiltonian system has been scaled so that its spectral range is comparable to that of the $L_B$ Hamiltonian, which broadens the banding structure for clarity.  The special role payed by the $\hat{H}^{(A)}$ Hamiltonian and its spectral gaps may be understood by considering by considering the microcanonical scrambling ansatz, as was described in section \ref{subsec: microcanonical ansatz - off diagonal}.

\bibliographystyle{unsrt}
\bibliography{paper}

\begin{thebibliography}{10}

\bibitem{Jensen:PhysRevLett.54.1879}
R.~V. Jensen and R.~Shankar.
\newblock Statistical behavior in deterministic quantum systems with few degrees of freedom.
\newblock {\em Phys. Rev. Lett.}, 54:1879--1882, Apr 1985.

\bibitem{Deutsch:PhysRevA.43.2046}
J.~M. Deutsch.
\newblock Quantum statistical mechanics in a closed system.
\newblock {\em Phys. Rev. A}, 43:2046--2049, Feb 1991.

\bibitem{Srednicki:PhysRevE.50.888}
Mark Srednicki.
\newblock Chaos and quantum thermalization.
\newblock {\em Phys. Rev. E}, 50:888--901, Aug 1994.

\bibitem{PhysRevE.87.012118}
R.~Steinigeweg, J.~Herbrych, and P.~Prelov\ifmmode~\check{s}\else \v{s}\fi{}ek.
\newblock Eigenstate thermalization within isolated spin-chain systems.
\newblock {\em Phys. Rev. E}, 87:012118, Jan 2013.

\bibitem{PhysRevE.89.042112}
W.~Beugeling, R.~Moessner, and Masudul Haque.
\newblock Finite-size scaling of eigenstate thermalization.
\newblock {\em Phys. Rev. E}, 89:042112, Apr 2014.

\bibitem{PhysRevE.90.052105}
Hyungwon Kim, Tatsuhiko~N. Ikeda, and David~A. Huse.
\newblock Testing whether all eigenstates obey the eigenstate thermalization hypothesis.
\newblock {\em Phys. Rev. E}, 90:052105, Nov 2014.

\bibitem{PhysRevE.89.062110}
E.~J. Torres-Herrera and Lea~F. Santos.
\newblock Local quenches with global effects in interacting quantum systems.
\newblock {\em Phys. Rev. E}, 89:062110, Jun 2014.

\bibitem{PhysRevE.93.032104}
Rubem Mondaini, Keith~R. Fratus, Mark Srednicki, and Marcos Rigol.
\newblock Eigenstate thermalization in the two-dimensional transverse field ising model.
\newblock {\em Phys. Rev. E}, 93:032104, Mar 2016.

\bibitem{PhysRevE.96.012157}
Rubem Mondaini and Marcos Rigol.
\newblock Eigenstate thermalization in the two-dimensional transverse field ising model. ii. off-diagonal matrix elements of observables.
\newblock {\em Phys. Rev. E}, 96:012157, Jul 2017.

\bibitem{PhysRevB.99.155130}
David Jansen, Jan Stolpp, Lev Vidmar, and Fabian Heidrich-Meisner.
\newblock Eigenstate thermalization and quantum chaos in the holstein polaron model.
\newblock {\em Phys. Rev. B}, 99:155130, Apr 2019.

\bibitem{PhysRevE.100.062134}
Tyler LeBlond, Krishnanand Mallayya, Lev Vidmar, and Marcos Rigol.
\newblock Entanglement and matrix elements of observables in interacting integrable systems.
\newblock {\em Phys. Rev. E}, 100:062134, Dec 2019.

\bibitem{PhysRevLett.125.070605}
Marlon Brenes, Tyler LeBlond, John Goold, and Marcos Rigol.
\newblock Eigenstate thermalization in a locally perturbed integrable system.
\newblock {\em Phys. Rev. Lett.}, 125:070605, Aug 2020.

\bibitem{Noh_2021}
Jae~Dong Noh.
\newblock Eigenstate thermalization hypothesis and eigenstate-to-eigenstate fluctuations.
\newblock {\em Physical Review E}, 103(1), January 2021.

\bibitem{Caceres:2024mec}
Elena C\'aceres, Stefan Eccles, Jason Pollack, and Sarah Racz.
\newblock {Generic ETH: Eigenstate Thermalization beyond the Microcanonical}.
\newblock 3 2024.

\bibitem{PhysRevB.109.045139}
Ding-Zu Wang, Hao Zhu, Jian Cui, Javier Arg\"uello-Luengo, Maciej Lewenstein, Guo-Feng Zhang, Piotr Sierant, and Shi-Ju Ran.
\newblock Eigenstate thermalization and its breakdown in quantum spin chains with inhomogeneous interactions.
\newblock {\em Phys. Rev. B}, 109:045139, Jan 2024.

\bibitem{Lin:2024vji}
Feng-Li Lin, Jhh-Jing Hong, and Ching-Yu Huang.
\newblock {Subsystem Thermalization Hypothesis in Quantum Spin Chains with Conserved Charges}.
\newblock 12 2024.

\bibitem{Lasek:2024ess}
Aleksander Lasek, Jae~Dong Noh, Jade LeSchack, and Nicole~Yunger Halpern.
\newblock {Numerical evidence for the non-Abelian eigenstate thermalization hypothesis}.
\newblock 12 2024.

\bibitem{DAlessio:2015qtq}
Luca D'Alessio, Yariv Kafri, Anatoli Polkovnikov, and Marcos Rigol.
\newblock {From quantum chaos and eigenstate thermalization to statistical mechanics and thermodynamics}.
\newblock {\em Adv. Phys.}, 65(3):239--362, 2016.

\bibitem{Deutsch:2018ulr}
Joshua~M. Deutsch.
\newblock {Eigenstate thermalization hypothesis}.
\newblock {\em Rept. Prog. Phys.}, 81(8):082001, 2018.

\bibitem{Reimann_2015}
Peter Reimann.
\newblock Eigenstate thermalization: Deutsch’s approach and beyond.
\newblock {\em New Journal of Physics}, 17(5):055025, may 2015.

\bibitem{Srednicki_1999}
Mark Srednicki.
\newblock The approach to thermal equilibrium in quantized chaotic systems.
\newblock {\em Journal of Physics A: Mathematical and General}, 32(7):1163–1175, January 1999.

\bibitem{Foini:2018sdb}
Laura Foini and Jorge Kurchan.
\newblock {Eigenstate thermalization hypothesis and out of time order correlators}.
\newblock {\em Phys. Rev. E}, 99(4):042139, 2019.

\bibitem{Chan:2018fsp}
Amos Chan, Andrea De~Luca, and J.~T. Chalker.
\newblock {Eigenstate Correlations, Thermalization and the Butterfly Effect}.
\newblock {\em Phys. Rev. Lett.}, 122(22):220601, 2019.

\bibitem{Dymarsky:2018ccu}
Anatoly Dymarsky.
\newblock {Bound on Eigenstate Thermalization from Transport}.
\newblock {\em Phys. Rev. Lett.}, 128(19):190601, 2022.

\bibitem{Belin:2021ryy}
Alexandre Belin, Jan de~Boer, and Diego Liska.
\newblock {Non-Gaussianities in the statistical distribution of heavy OPE coefficients and wormholes}.
\newblock {\em JHEP}, 06:116, 2022.

\bibitem{PhysRevLett.111.050403}
Ehsan Khatami, Guido Pupillo, Mark Srednicki, and Marcos Rigol.
\newblock Fluctuation-dissipation theorem in an isolated system of quantum dipolar bosons after a quench.
\newblock {\em Phys. Rev. Lett.}, 111:050403, Jul 2013.

\bibitem{Murthy:2019fgs}
Chaitanya Murthy and Mark Srednicki.
\newblock {Bounds on chaos from the eigenstate thermalization hypothesis}.
\newblock {\em Phys. Rev. Lett.}, 123(23):230606, 2019.

\bibitem{Sorokhaibam:2022tgq}
Nilakash Sorokhaibam.
\newblock {Quantum chaos and the arrow of time}.
\newblock 12 2022.

\bibitem{Sorokhaibam:2024fcv}
Nilakash Sorokhaibam.
\newblock {A measure of chaos from eigenstate thermalization hypothesis}.
\newblock 1 2024.

\bibitem{PhysRevLett.105.250401}
Giulio Biroli, Corinna Kollath, and Andreas~M. L\"auchli.
\newblock Effect of rare fluctuations on the thermalization of isolated quantum systems.
\newblock {\em Phys. Rev. Lett.}, 105:250401, Dec 2010.

\bibitem{Kawamoto:2024vzd}
Taishi Kawamoto.
\newblock {A Strategy for Proving the Strong Eigenstate Thermalization Hypothesis : Chaotic Systems and Holography}.
\newblock 11 2024.

\bibitem{vonNeumann_1929}
J.~von Neumann.
\newblock Proof of the ergodic theorem and the h-theorem in quantum mechanics.
\newblock {\em The European Physical Journal H}, 35(2):201--237, 2010.

\bibitem{von_Neumann_2010}
J.~von Neumann.
\newblock Proof of the ergodic theorem and the h-theorem in quantum mechanics: Translation of: Beweis des ergodensatzes und des h-theorems in der neuen mechanik.
\newblock {\em The European Physical Journal H}, 35(2):201–237, September 2010.

\bibitem{Goldstein_2010}
S.~Goldstein, J.~L. Lebowitz, R.~Tumulka, and N.~Zanghì.
\newblock Long-time behavior of macroscopic quantum systems: Commentary accompanying the english translation of john von neumann’s 1929 article on the quantum ergodic theorem.
\newblock {\em The European Physical Journal H}, 35(2):173–200, September 2010.

\bibitem{PhysRevLett.115.010403}
Peter Reimann.
\newblock Generalization of von neumann's approach to thermalization.
\newblock {\em Phys. Rev. Lett.}, 115:010403, Jul 2015.

\bibitem{Ishii_2019}
Takashi Ishii and Takashi Mori.
\newblock Strong eigenstate thermalization within a generalized shell in noninteracting integrable systems.
\newblock {\em Physical Review E}, 100(1), July 2019.

\bibitem{Helbig:2024lvx}
Tobias Helbig, Tobias Hofmann, Ronny Thomale, and Martin Greiter.
\newblock {Theory of Eigenstate Thermalisation}.
\newblock 6 2024.

\bibitem{Jafferis:2022uhu}
Daniel~Louis Jafferis, David~K. Kolchmeyer, Baur Mukhametzhanov, and Julian Sonner.
\newblock {Matrix Models for Eigenstate Thermalization}.
\newblock {\em Phys. Rev. X}, 13(3):031033, 2023.

\bibitem{jafferis2023jtgravitymattergeneralized}
Daniel~Louis Jafferis, David~K. Kolchmeyer, Baur Mukhametzhanov, and Julian Sonner.
\newblock Jt gravity with matter, generalized eth, and random matrices, 2023.

\bibitem{deBoer:2023vsm}
Jan de~Boer, Diego Liska, Boris Post, and Martin Sasieta.
\newblock {A principle of maximum ignorance for semiclassical gravity}.
\newblock {\em JHEP}, 2024:003, 2024.

\bibitem{RN7}
R.~Balian.
\newblock Random matrices and information theory.
\newblock {\em Il Nuovo Cimento B (1965-1970)}, 57(1):183--193, 1968.

\bibitem{PhysRev.106.620}
E.~T. Jaynes.
\newblock Information theory and statistical mechanics.
\newblock {\em Phys. Rev.}, 106:620--630, May 1957.

\bibitem{PhysRevLett.111.127201}
Maksym Serbyn, Z.~Papi\ifmmode~\acute{c}\else \'{c}\fi{}, and Dmitry~A. Abanin.
\newblock Local conservation laws and the structure of the many-body localized states.
\newblock {\em Phys. Rev. Lett.}, 111:127201, Sep 2013.

\bibitem{PhysRevB.90.174202}
David~A. Huse, Rahul Nandkishore, and Vadim Oganesyan.
\newblock Phenomenology of fully many-body-localized systems.
\newblock {\em Phys. Rev. B}, 90:174202, Nov 2014.

\bibitem{annurev:/content/journals/10.1146/annurev-conmatphys-031214-014726}
Rahul Nandkishore and David~A. Huse.
\newblock Many-body localization and thermalization in quantum statistical mechanics.
\newblock {\em Annual Review of Condensed Matter Physics}, 6(Volume 6, 2015):15--38, 2015.

\bibitem{RN8}
John~Z. Imbrie.
\newblock On many-body localization for quantum spin chains.
\newblock {\em Journal of Statistical Physics}, 163(5):998--1048, 2016.

\bibitem{RevModPhys.91.021001}
Dmitry~A. Abanin, Ehud Altman, Immanuel Bloch, and Maksym Serbyn.
\newblock Colloquium: Many-body localization, thermalization, and entanglement.
\newblock {\em Rev. Mod. Phys.}, 91:021001, May 2019.

\bibitem{Rigol_Dunjko_Yurovsky_PhysRevLett.98.050405}
Marcos Rigol, Vanja Dunjko, Vladimir Yurovsky, and Maxim Olshanii.
\newblock Relaxation in a completely integrable many-body quantum system: An ab initio study of the dynamics of the highly excited states of 1d lattice hard-core bosons.
\newblock {\em Phys. Rev. Lett.}, 98:050405, Feb 2007.

\bibitem{PhysRevLett.106.140405}
Amy~C. Cassidy, Charles~W. Clark, and Marcos Rigol.
\newblock Generalized thermalization in an integrable lattice system.
\newblock {\em Phys. Rev. Lett.}, 106:140405, Apr 2011.

\bibitem{PhysRevA.87.063637}
Kai He, Lea~F. Santos, Tod~M. Wright, and Marcos Rigol.
\newblock Single-particle and many-body analyses of a quasiperiodic integrable system after a quench.
\newblock {\em Phys. Rev. A}, 87:063637, Jun 2013.

\bibitem{Vidmar_2016}
Lev Vidmar and Marcos Rigol.
\newblock Generalized gibbs ensemble in integrable lattice models.
\newblock {\em Journal of Statistical Mechanics: Theory and Experiment}, 2016(6):064007, jun 2016.

\bibitem{santos2010onset}
Lea~F Santos and Marcos Rigol.
\newblock Onset of quantum chaos in one-dimensional bosonic and fermionic systems and its relation to thermalization.
\newblock {\em Physical Review E}, 81(3):036206, 2010.

\bibitem{PhysRevLett.130.140402}
Chaitanya Murthy, Arman Babakhani, Fernando Iniguez, Mark Srednicki, and Nicole Yunger~Halpern.
\newblock Non-abelian eigenstate thermalization hypothesis.
\newblock {\em Phys. Rev. Lett.}, 130:140402, Apr 2023.

\bibitem{Meh2004}
Madan~Lal Mehta.
\newblock {\em Random Matrices}.
\newblock 3rd edition, 2004.

\bibitem{Cotler:2017abq}
Jordan~S. Cotler, Geoffrey~R. Penington, and Daniel~H. Ranard.
\newblock {Locality from the Spectrum}.
\newblock {\em Commun. Math. Phys.}, 368(3):1267--1296, 2019.

\bibitem{Goldstein_2006}
Sheldon Goldstein, Joel~L. Lebowitz, Roderich Tumulka, and Nino Zangh{\`{\i}}.
\newblock Canonical typicality.
\newblock {\em Physical Review Letters}, 96(5), feb 2006.

\bibitem{Banuls:2011vuw}
M.~C. Ba\~nuls, J.~I. Cirac, and M.~B. Hastings.
\newblock {Strong and Weak Thermalization of Infinite Nonintegrable Quantum Systems}.
\newblock {\em Phys. Rev. Lett.}, 106(5):050405, 2011.

\bibitem{Shenker:2013pqa}
Stephen~H. Shenker and Douglas Stanford.
\newblock {Black holes and the butterfly effect}.
\newblock {\em JHEP}, 03:067, 2014.

\bibitem{Hosur:2015ylk}
Pavan Hosur, Xiao-Liang Qi, Daniel~A. Roberts, and Beni Yoshida.
\newblock {Chaos in quantum channels}.
\newblock {\em JHEP}, 02:004, 2016.

\bibitem{Couch:2019zni}
Josiah Couch, Stefan Eccles, Phuc Nguyen, Brian Swingle, and Shenglong Xu.
\newblock {Speed of quantum information spreading in chaotic systems}.
\newblock {\em Phys. Rev. B}, 102(4):045114, 2020.

\bibitem{Eccles:2021zum}
Stefan Eccles, Willy Fischler, Tyler Guglielmo, Juan~F. Pedraza, and Sarah Racz.
\newblock {Speeding up the spread of quantum information in chaotic systems}.
\newblock {\em JHEP}, 12:019, 2021.

\bibitem{Bianchi_2024}
Eugenio Bianchi, Pietro Dona, and Rishabh Kumar.
\newblock Non-abelian symmetry-resolved entanglement entropy.
\newblock {\em SciPost Physics}, 17(5), November 2024.

\end{thebibliography}
\end{document}